\documentclass[twocolumn,showpacs,preprintnumbers,amsmath,amssymb]{revtex4}

\usepackage{epsfig}
\usepackage{graphicx}                          % Include figure files
\usepackage{dcolumn}                          % Align table columns on decimal point
\usepackage{bm}                                   % bold math
\usepackage{subfigure}

\newcommand{\beq}{\begin{equation}}
\newcommand{\eeq}{\end{equation}}
\newcommand{\beqa}{\begin{eqnarray}}
\newcommand{\eeqa}{\end{eqnarray}}

\newcommand{\la}{\langle}
\newcommand{\ra}{\rangle}
\newcommand{\ket}[1]{|#1\rangle}                                % Ket
\newcommand{\bra}[1]{\langle#1|}                                % Bra
\newcommand{\scpr}[2]{\langle#1|#2\rangle}                % Scalar Product
\newcommand{\matel}[3]{\langle#1|#2|#3\rangle}         % Matrix Element

\def\jmo#1{{ J.\ Mod.\ Opt.} {\bf#1}}
\def\jpa#1{{ J.\ Phys.\ A} {\bf#1}}

\def\pra#1{{ Phys.\ Rev. A\/} {\bf#1}}
\def\prb#1{{ Phys.\ Rev. B\/} {\bf#1}}

\def\prl#1{{ Phys.\ Rev.\ Lett.} {\bf#1}}

\begin{document}

\title{Tavis-Cummings model beyond the rotating wave approximation: Quasi-degenerate qubits}

\author{S. Agarwal, S.M. Hashemi Rafsanjani and J.H. Eberly}

\affiliation{ Rochester Theory Center and the Department of Physics
\& Astronomy\\
University of Rochester, Rochester, New York 14627}

\email{shantanu@pas.rochester.edu}

%\author{.........}
%\email{......@pas.rochester.edu}
%\author{ll}
%\author{}
%\preprint{}

\date{\today}

\begin{abstract}
The Tavis-Cummings model for more than one qubit interacting with a common oscillator mode is extended beyond the rotating wave approximation (RWA). We explore the parameter regime in which the frequencies of the qubits are much smaller than the oscillator frequency and the coupling strength is allowed to be ultra-strong. The application of the adiabatic approximation, introduced by Irish, et al. (Phys. Rev. B \textbf{72}, 195410 (2005)), for a single qubit system is extended to the multi-qubit case. For a two-qubit system, we identify three-state manifolds of close-lying dressed energy levels and obtain results for the dynamics of intra-manifold transitions that are incompatible with results from the familiar regime of the RWA. We exhibit features of two-qubit dynamics that are different from the single qubit case, including calculations of qubit-qubit entanglement. Both number state and coherent state preparations are considered, and we derive analytical formulas that simplify the interpretation of numerical calculations. Expressions for individual collapse and revival signals of both population and entanglement are derived. 
\end{abstract}

% PACS
%\keywords{Suggested keywords}%Use showkeys class option if keyword
%display desired

\maketitle

%===================================================

\section{Introduction}\label{introduction}

%===================================================

Two level systems that interact with a harmonic oscillator can model many physical phenomena, such as nuclear spins interacting with magnetic field \cite{Rabi}, atoms interacting 
with electromagnetic field \cite{JC, Allen-Eberly}, electrons coupled to a phonon mode of a crystal lattice \cite{Holstein}, super-conducting qubits interacting with a nano-mechanical resonator \cite{Irish-Schwab, Schwab-Roukes}, a transmission line resonator \cite{Blais-etal, Wallraff-etal}, or an LC circuit \cite{Chiorescu-etal, Johansson-etal}, etc. The dynamics of all such systems is governed by the Rabi Hamiltonian \cite{Rabi}: 
\beq\label{RabiHam}
\hat{H}^{Rabi} = \frac{\hbar\omega_{0}}{2}\hat{\sigma}_{z} + \hbar\omega\hat{a}^{\dagger}\hat{a}
+ {\hbar\omega\frac{\beta}{2}}(\hat{a}+\hat{a}^{\dagger})(\hat{\sigma}_{+} +\hat{\sigma}_{-}),
\eeq
where the $\hat\sigma_z$ and $\hat\sigma_+ +\hat\sigma_-=\hat\sigma_x$ are the usual Pauli matrices in the Hilbert space of the qubit and $\hat a^\dag$ and $\hat a$ refer to the creation and annihilation operators of an interacting mode of a harmonic oscillator. Although studied extensively since it was first introduced in the context of nuclear magnetic spin resonance, analytical solutions for the eigenvalues and eigenfunctions of the Rabi Hamiltonian still do not exist.

In physical situations where the qubits are nearly resonant with the oscillator and the coupling strengths between the qubits and the oscillator are much smaller than the qubit and the oscillator frequencies, it is a good approximation to drop the counter rotating terms: $\hat a^\dag\hat\sigma_+$ and $\hat a\hat\sigma_-$, from (\ref{RabiHam}) to obtain the so-called Jaynes-Cummings (JC) model with the Hamiltonian \cite{JC}:
\beq\label{JCHam}
\hat{H}^{JC} = \frac{\hbar\omega_{0}}{2}\hat{\sigma}_{z} + \hbar\omega\hat{a}^{\dagger}\hat{a} +  {\hbar\omega\frac{\beta}{2}}(\hat a\hat{\sigma}_{+} + \hat{\sigma}_{-}\hat a^\dag).
\eeq
Under this approximation, called the rotating wave approximation (RWA), the dynamics of the system can be obtained in closed form \cite{JC,Allen-Eberly}. 

A generalization of the JC model, called the Tavis-Cummings (TC) model, was introduced in the context of quantum optics to describe the collective behavior of multiple atomic dipoles interacting with an electromagnetic field mode \cite{Dicke, Tavis-Cummings-1, Tavis-Cummings-2}. The TC model has gained renewed interest as it can be used to implement quantum information protocols with the oscillator transferring information coherently between qubits \cite{Leek}. Intrinsically  multi-qubit properties such as quantum entanglement can be explored with the TC model in a variety of ways, employing various entanglement measures such as concurrence for mixed-state pairs of qubits \cite{Wootters}, quantum negativity for slightly larger systems \cite{Peres}, and Schmidt weights for bipartitions of arbitrarily dimensioned pure multi-qubit states \cite{Grobe}.

%===================================================

\section{Multi-Qubit Breakdown of the RWA}\label{Breakdown}

%===================================================

\begin{figure}[!t]
\includegraphics[width= 6cm]{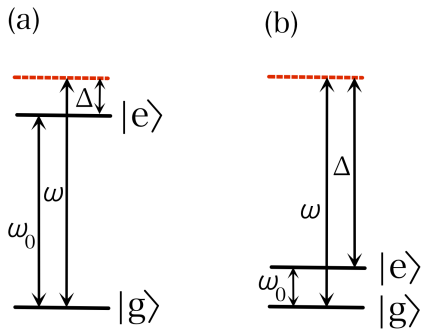} 
\caption{\label{f.model} Diagrams showing energy-level configurations: (a)  compatible with the RWA, $\Delta \ll \omega_0 \approx \omega$; (b) incompatible with the RWA, $\Delta \gg \omega_0 \ll \omega$. The states $\ket{e}$ and $\ket{g}$ are the eigenfunctions of $\hat{\sigma}_z$: $\hat{\sigma}_z\ket{e}=\ket{e}$ and $\hat{\sigma}_z\ket{g}=-\ket{g}$.} 
\end{figure}

With recent advances in the area of circuit QED, it is now possible to engineer systems for which the qubits are coupled to the oscillator so strongly, or are so far detuned from the oscillator, that the RWA cannot be used to describe the system's evolution correctly \cite{Niemczyk-etal, Forn-etal, Fedorov-etal}. 
The parameter regime for which the coupling strength is strong enough to invalidate the RWA is called the ultra-strong coupling regime \cite{Irish-05, Irish-07, Hausinger-10, Devoret, Bourassa, Ashhab, Casanova}. 
Niemczyk, et al. \cite{Niemczyk-etal} and Forn-D\'{i}az, et al. \cite{Forn-etal} have been able to experimentally achieve ultra-strong coupling strengths and have demonstrated the breakdown of the RWA.  Motivated by these experimental developments and the importance of understanding collective quantum behavior, we investigate a two-qubit TC model beyond the validity regime of RWA. 
The regime of parameters we will be concerned with is the regime where the qubits are quasi-degenerate, i.e., with frequencies much smaller than the oscillator frequency, $\omega_0 \ll \omega$, while the coupling between the qubits and the oscillator is allowed to be an appreciable fraction of the oscillator frequency. In this parameter regime, the dynamics of the system can neither be correctly described under the RWA, nor can the effects of the counter rotating terms be taken as a perturbative correction to the dynamics predicted within the RWA by including higher powers of $\beta$. 
For illustration, systems are shown in Fig. \ref{f.model} for which the RWA is valid, or breaks down, because the condition $\omega_{0} \approx \omega$ is valid, or is violated. The regime that we will be interested in, for which $\omega_0\ll\omega$, is shown on the right. 

Prior numerical work by Irish has been directed to the dynamics of a single quasi-degenerate qubit interacting with an oscillator in the ultra-strong coupling regime, and carried out  by developing an adiabatic approximation \cite{Irish-05}, with an extension to a generalized RWA  \cite{Irish-07}, and also by Hausinger, et al., by using van Vleck perturbation theory \cite{Hausinger-10}. The adiabatic approximation and van Vleck perturbation theory were shown to work best for small qubit frequencies and high coupling strengths. The adiabatic approximation was shown to fail in the regime where the JC model works well, i.e., when the qubit is resonant with the oscillator and the coupling is small. This gap between the regime of validity of the adiabatic approximation and the regime of validity of JC model was bridged by the generalized RWA \cite{Irish-07}, which works well in both regimes. 

Here, within the adiabatic approximation, we extend the examination to the two-qubit case. Qualitative differences between the single-qubit and the multi-qubit cases are highlighted. In particular, we study the collapse and revival of joint properties of both the qubits. Entanglement properties of the system are investigated and it is shown that the entanglement between the qubits also exhibits collapse and revival. We derive what we believe are the first analytic expressions for the individual revival signals beyond the RWA, as well as analytic expression for the collapse and revival dynamics of entanglement. In the quasi-degenerate regime, the invalidity of the RWA in predicting the dynamical evolution will clearly be demonstrated in Sec. \ref{s.collapse_rev} (see Figs. \ref{f.collapse_revival_double} and \ref{f.collapse_revival_single}).

We begin with a generalization of (\ref{RabiHam}) in which the $\hat\sigma$ operators are replaced by two-qubit counterparts \cite{Tavis-Cummings-1}:
\beq\label{Hamiltonian}
\hat{H}=\hbar\omega_{0}\hat{S}_{z} + \hbar\omega \hat{a}^{\dagger}\hat{a}
+ \hbar\omega\beta(\hat{a}+\hat{a}^{\dagger})\hat{S}_{x},
\eeq
where 
\beq
\hat{S}_{z} = \frac{1}{2}(\hat{\sigma}_{z}^{(1)}+\hat{\sigma}_{z}^{(2)}),\ {\rm and}\ 
\hat{S}_{x} = \frac{1}{2}(\hat{\sigma}_{x}^{(1)}+\hat{\sigma}_{x}^{(2)}).
\eeq
In experiments dealing with artificial qubits, such as Cooper pair boxes, it is possible to bias the qubits, which results in an additional term in the Hamiltonian:
\beq
H_{bias}=\hbar\epsilon\hat{S}_{x},
\eeq
where $\epsilon$ is called the static bias. Taking finite bias into account, $\epsilon\neq0$, an analysis of the dynamics of a single qubit interacting with a harmonic oscillator beyond the RWA in the ultra-strong coupling regime was done in \cite{Hausinger-10, Ashhab}. Here we assume that $\epsilon=0$.

%===================================================

\section{Informal Analysis}\label{Informal}

%===================================================
Before proceeding with a detailed treatment, we note that an informal approach to the Hamiltonian (\ref{Hamiltonian}) is possible, and can be helpful in interpreting further analysis. The disparity in time scales signaled by the inequality $\omega_0 \ll \omega$ governs new effects that will occur. To see this, we let $\omega_0$ be sufficiently small as to be negligible, thus removing the $\hat{S}_z$ from any role in $\hat H$. Then $\hat S_x$ becomes constant, say $\hat S_x(0)$. The Heisenberg equation for the response of the oscillator amplitude $\hat a$ becomes trivial, with the solution
\beq \label{a-degensoln}
\hat a(t) + \beta \hat S_x(0) = (\hat a(0) + \beta \hat S_x(0))e^{-i\omega t},
\eeq
which is easily interpreted in the expected-value sense: the evolution of $\la\hat a \ra$ is sinusoidal at frequency $\omega$ and centered at $-\beta\la\hat S_x(0)\ra$. 

Of course, $\hat S_x$ is not constant if $\omega_0\neq 0$. As an operator, it evolves in time. Its evolution is determined by the commutator with $\hat H$, and this leads to the three coupled Bloch-type equations:
\beqa \label{BlochEqns}
d\hat S_x/dt &=& -\omega_0 \hat S_y, \\
d\hat S_y/dt &=& \omega_0 \hat S_x -\beta\omega (\hat a + \hat a^\dag)\hat S_z, \\
d\hat S_z/dt &=& \beta\omega(\hat a + \hat a^\dag) \hat S_y. 
\eeqa
Under realistic current laboratory conditions $\beta \ll 1$, so unless the oscillator amplitude is very great the main spin motion is a slow precession of $\hat S_x$ and $\hat S_y$ at frequency $\omega_0$, with small and very rapid oscillations at frequency $\omega$ arising from the $\beta(a + a^\dag)$ terms.

Two comments are obvious at this level of analysis. First, since $\hat S_x$ changes nearly periodically on the time scale $\sim 2\pi/\omega_0$, we expect the center of oscillator motion to follow these slow changes back and forth. Second, the rapid oscillations around the slow precession are of both signs $\pm \omega$, so they contain the effect of the counter-rotating terms omitted by the JC model. 
It should be noted that this informal analysis is not specific to any particular number of qubits and all the comments of this section are equally applicable to an $K$-qubit system.

%===================================================

\section{Spectrum of $\hat{H}$}\label{s.Spectrum}

%===================================================

We first find the eigenspectrum of $\hat{H}$ when $\omega_0=0$.
The Hamiltonian without the RWA then takes the form 
\beq\label{e.H_o}
\hat{H}_{0}=\hbar\omega \hat{a}^{\dagger}\hat{a}
+ \hbar\beta\omega(\hat{a}+\hat{a}^{\dagger})\hat{S}_x.
\eeq
The eigenstates and eigenvalues of $\hat{H}_0$ satisfy the eigenvalue equation:
\beq\label{e.Degen_Eigen1}
\hbar\omega \left[\hat{a}^{\dagger}\hat{a}
+\beta(\hat{a}+\hat{a}^{\dagger})\hat{S}_x\right]\ket{\Phi}=E\ket{\Phi}.
\eeq
The eigenstates $\ket{\Phi}$ will be products of qubits and oscillator states, and take the form 
\beq
\ket{\Phi} = \ket{j,m}\ket{\phi_{m}}.
\eeq 
Here $\ket{j,m}$ are the eigenstates of 
$\hat{S}_x$ and $\ket{\phi_{m}}$ are the oscillator eigenstates found from 
$\hat{H}_0$ by replacing $\hat{S}_x$ by its eigenvalue corresponding to $\ket{j,m}$ \cite{Irish-05}. 

The four eigenstates of $\hat{S}_x$ are:
\beq
\ket{j,m}=\ket{1,\pm1}\mbox{, }\ket{1,0}\mbox{ and }\ket{0,0},
\eeq
with eigenvalues $m$.
In terms of the simultaneous eigenstates of $\hat{\sigma}_x^{(1)}$ and $\hat{\sigma}_x^{(2)}$, 
$\hat{\sigma}_x^{(i)}\ket{\pm}=\pm\ket{\pm}$, the states $\ket{j,m}$ can be written as:
\begin{equation}\label{e.unit_trans}
 \begin{pmatrix} \ket{1,1} \\ \ket{1,0} \\ \ket{0,0} \\ \ket{1,-1}\end{pmatrix}=
 \begin{pmatrix} 1 & 0 & 0 & 0 \\ 
                 0 & 1/\sqrt{2} & 1/\sqrt{2} & 0 \\
                 0 & 1/\sqrt{2} & -1/\sqrt{2} & 0 \\
                 0 & 0 &  0 & 1 \\
 \end{pmatrix}
 \begin{pmatrix} \ket{+,+} \\ \ket{+,-} \\ \ket{-,+} \\ \ket{-,-}\end{pmatrix}.
\end{equation}

Having found $\ket{j,m}$, let us now find $\ket{\phi_{m}}$ that satisfy the eigenvalue equation: 
\beq\label{e.Degen_Eigen2}
\hbar\omega \left[\hat{a}^{\dagger}\hat{a}
+ m\beta(\hat{a}+\hat{a}^{\dagger})\right]\ket{\phi_{m}} = E\ket{\phi_{m}}.
\eeq  
We denote $m\beta$ by $\beta_m$, which we take real. Then by completing the square in (\ref{e.Degen_Eigen2}), we get a new number operator equation:
\beqa\label{e.Degen_Eigen3}
(\hat{a}^{\dagger}+\beta_m)(\hat{a}+\beta_m)\ket{\phi_m}
&=&\left({E}/{\hbar\omega}+\beta_m^2\right)\ket{\phi_m} \nonumber \\
= N|\phi_m\ra,&&  N = 0, 1, \dots.
\eeqa 
Using the displacement operator, $\hat{D}(\alpha) = \exp[\alpha(\hat{a}^{\dagger}-\hat{a})]$ (for real $\alpha$), we can write the expression on the left side of (\ref{e.Degen_Eigen3}) as $\hat{D}^\dag(\beta_m)\hat{a}^{\dagger}\hat{a}\hat{D}(\beta_m)\ket{\phi_m}$. Then multiplication of this by $D(\beta_m)$ converts (\ref{e.Degen_Eigen3}) into
\beqa\label{e.Degen_Eigen4}
\hat{a}^{\dagger}\hat{a}\Big(\hat{D}(\beta_m)\ket{\phi_m}\Big) = N\Big(\hat{D}(\beta_m)\ket{\phi_m}\Big),
\eeqa
which shows that the original oscillator and its displaced counterpart have the same eigenvalues, and relates their eigenstates as
\beqa\label{e.dis_Fock}
D(\beta_m)|\phi_m^N\ra &=& |N\ra \quad {\rm or}\nonumber\\
|\phi_m^N\ra &=& D(-\beta_m)|N\ra \equiv |N_m\ra. 
\eeqa
Thus, finally, the joint qubit-oscillator eigenstates are of the form:
\beq
|\Phi \ra \to |\Phi_{j,m,N} \ra = |j,m\ra\ |N_m\ra,
\eeq
and the energy $E$ in (\ref{e.Degen_Eigen2}) takes values:
\beq \label{e.energy_om0}
E_{N,m} = \hbar\omega(N - \beta_m^2).
\eeq

Thus, we see that depending upon the state of the qubits, determined by $\ket{j,m}$, we have four harmonic oscillator potential wells in $x-p$ phase space, where $\hat x = (\hat a^\dag + \hat a)/\sqrt 2$ and  $\hat p = i(\hat a^\dag - \hat a)/\sqrt 2$.  These potential wells have their equilibrium positions displaced by an amount proportional to $2m\beta$. For $m = 0$, the oscillator potentials are not displaced, whereas for $m=\pm1$, they are displaced in equal and opposite directions. A very important thing to note from (\ref{e.energy_om0}) is that the eigenstates with the same value of $N$ are not degenerate, e.g., the states $\ket{1,0}\ket{N_0}$ and $\ket{1,1}\ket{N_1}$ differ in energy by $\hbar\omega\beta^2$. For contrast, in the single qubit case, when $\omega_0=0$, the eigenstates of the Hamiltonian with the same value of $N$ remain degenerate irrespective of the value of $\beta$ \cite{Irish-05}.

\begin{figure}
\includegraphics[width= 8cm]{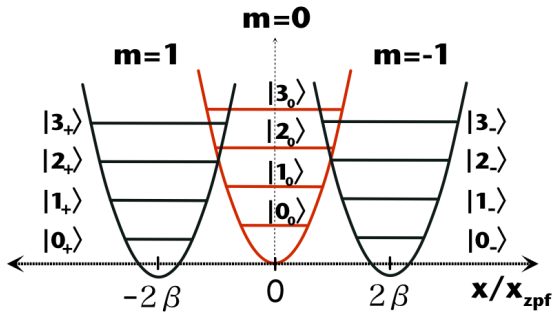}
\caption{The three potential wells corresponding to the states $\ket{1,1}\ket{N_1}$ (left), $\ket{1,0}\ket{N_0}$ (middle) and $\ket{1,-1}\ket{N_{-1}}$ (right). The factor $\Delta X_{zp}$ is the zero point fluctuation of a harmonic oscillator. For an oscillator of mass $M$ and frequency $\omega$ the zero point fluctuation is given by $\Delta X_{zp}=\sqrt{\hbar/2M\omega}$.} \label{f.tavis}
\end{figure}

The three potential wells corresponding to the states $\ket{1,m}\ket{N_m}$ are schematically shown in Fig. \ref{f.tavis}. The displacement of the equilibrium position of the potential wells and the relative lowering of the energy levels for $m=\pm1$ states is evident from the figure.

One may say that because of its coupling to the qubits the original oscillator is not really the ``effective" oscillator, with the consequence that a definite number of its excitations does not correspond to a definite number of the effective excitations, and vice versa. This is the nature of the displacement operation. In a discussion of two level systems interacting with a harmonic oscillator beyond the RWA, the use of a displaced harmonic oscillator basis was first used by Schweber \cite{Schweber-67}. The displaced oscillator states have the properties:
\beqa
\scpr{N_{m}}{N'_{m}} &=& \delta_{N,N'},\nonumber\\
\scpr{N_{m}}{N'_{m'}} &\neq& 0,
\eeqa
and in particular
\beq
\scpr{N_{1}}{N_{0}} = e^{-\beta^2/2}L_{N}(\beta^2),
\eeq
where $L_{N}(x)$ is a Laguerre polynomial.
The non-orthogonality condition, $\scpr{N_{m}}{N'_{m'}}\neq0$, plays an important role in subsequent analysis.

%===================================================
Next, we extend the discussion to examine the eigen-spectrum of $\hat{H}$ when $\omega_0\neq 0$. Using the basis $\ket{j,m}\ket{N_m}$ we now look for the eigenstates and eigenvalues of $\hat{H}$ when $\omega_0\neq0$. 
We note that because $\ket{0,0}$ is a simultaneous eigenstate of $\hat{S}_{z}$ and $\hat{S}_{x}$:
\beqa
\hat{S}_{z}\ket{0,0}&=&0,\nonumber\\
\hat{S}_{x}\ket{0,0}&=&0,
\eeqa
the states $\ket{0,0}\ket{N_0}$ (for any $N$) are eigenstates of $\hat{H}$ with eigenvalues $E_{N,0}=\hbar N\omega$, even when $\omega_0\neq0$. This allows one to find the exact evolution of the system in the projected Hilbert space spanned by the states $\ket{0,0}\ket{N_0}$.
However, finding the evolution of a state spanned by $\ket{1,m}\ket{N_m}$ is a challenge because the states $\ket{1,m}\ket{N_{m}}$ are not simultaneous eigenstates of $\hat{S}_{z}$ and $\hat{S}_{x}$. We now look at the Hamiltonian that is spanned by $\ket{1,m}\ket{N_m}$ states only. We assume that even though $\omega_0\neq0$, it is still small compared to the frequency of the oscillator as a result of which one can treat the term $\hbar\omega_0\hat{S}_{z}$ as a perturbation to the energy spectrum found for the case when $\omega_0=0$. In particular, we will restrict our analysis to the regime: $\omega_0\leq 0.25\omega$, which we label as the quasi-degenerate regime.

\begin{figure}[htb]
	\includegraphics[width=5cm]{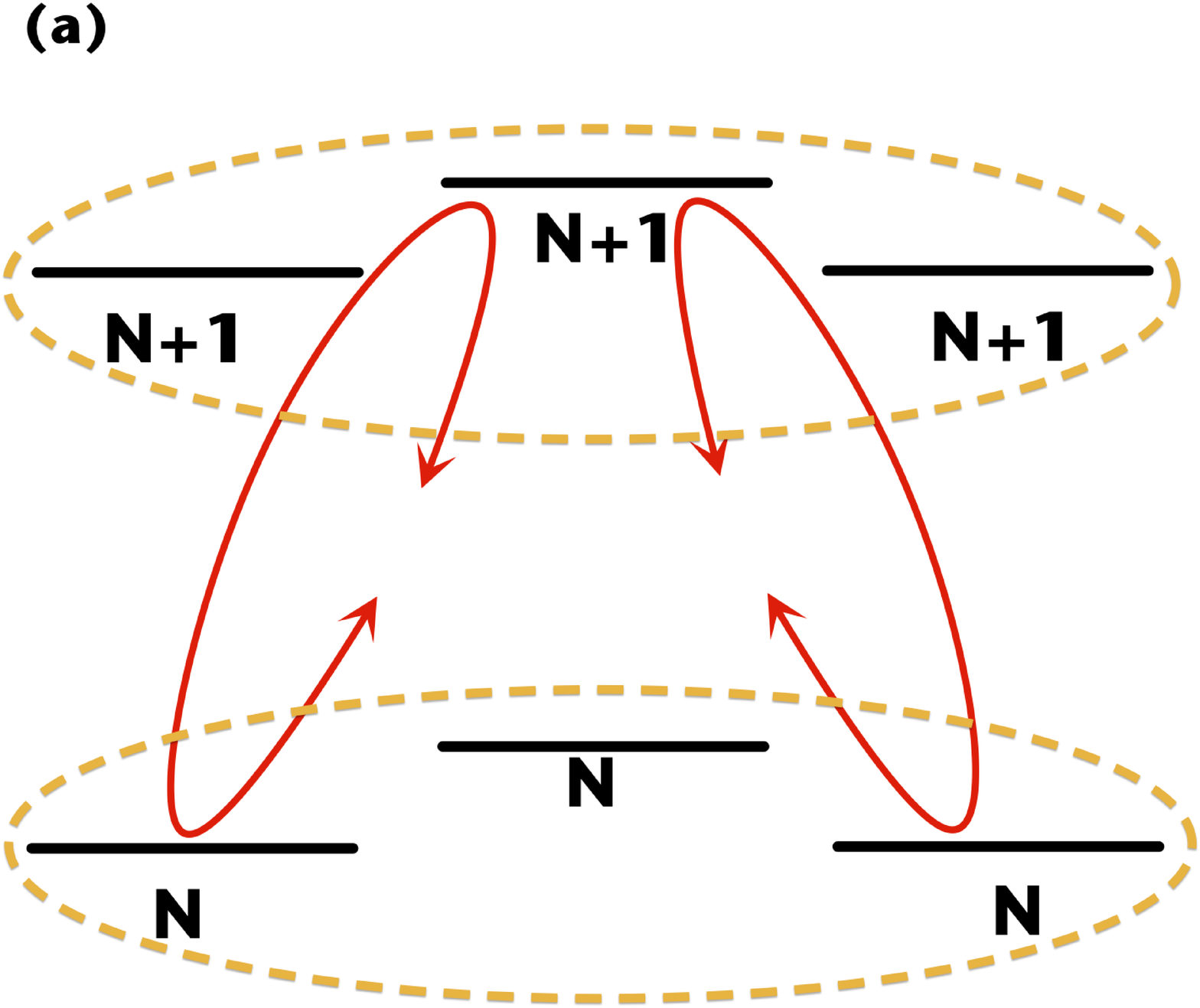}
	\includegraphics[width=5cm]{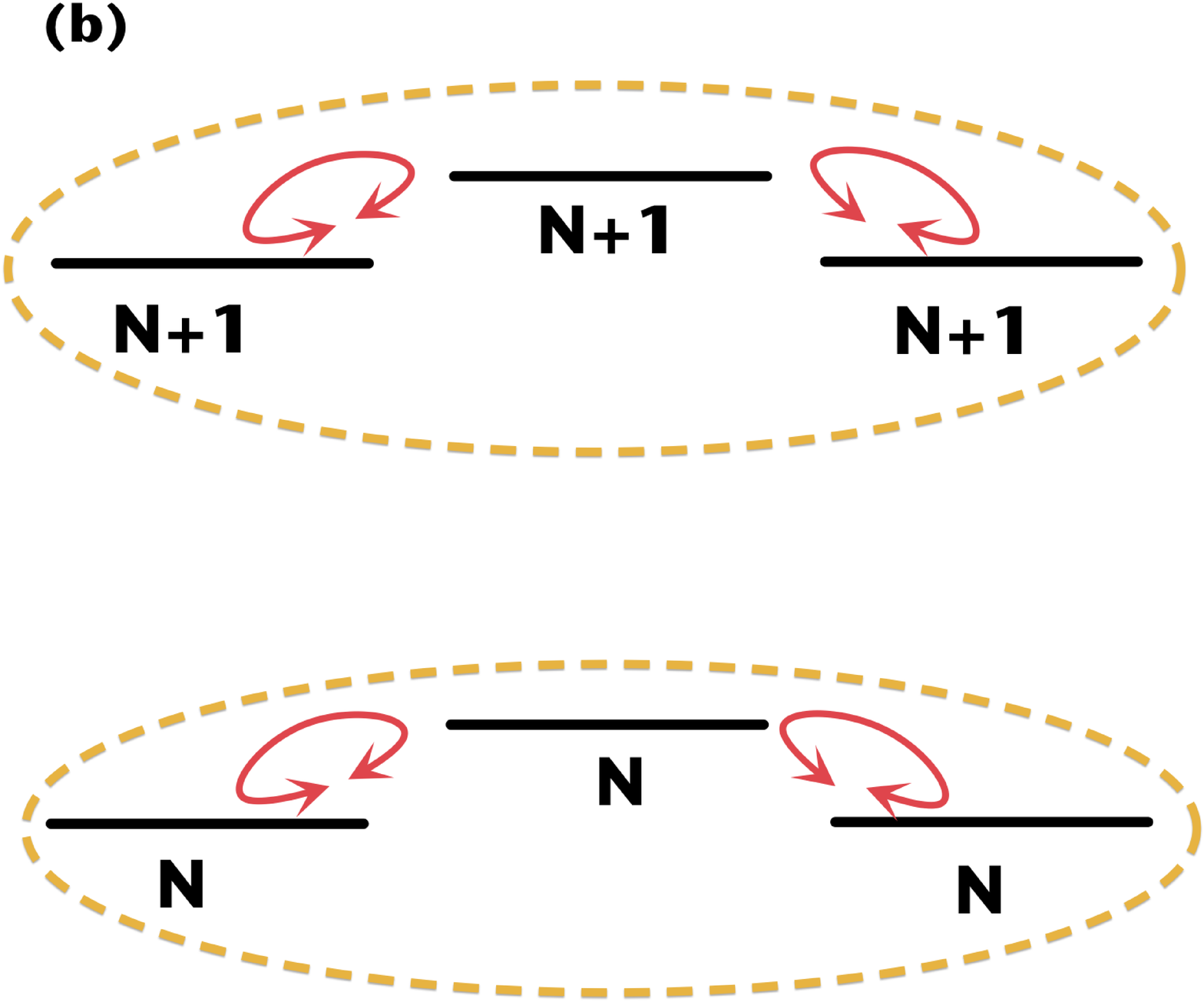}
	\caption{(a) Transitions induced by $\hbar\omega_0\hat{S}_z$ between states of different manifolds. (b) Transitions induced by $\hbar\omega_0\hat{S}_z$ between states of the same manifold.}
	\label{f.transitions}
\end{figure}

We start by noticing from (\ref{e.energy_om0}) that when $\omega_0=0$ and $\beta^2$ is close to an integer, say \textit{p}, the three states: $\ket{1,0}\ket{N_0}$ and $\ket{1,\pm1}\ket{(N+\textit{p})_{\pm1}}$, are grouped together in energy and are nearly degenerate. In what follows, we will not be concerned with very high values of $|\beta|$, but explore the regime that is experimentally achievable with current technology or is likely to be realizable within the near future. For this reason, we restrict our analysis to the regime where $|\beta| \leq 0.25$, which is strong enough to invalidate the RWA, i.e., which lies in the ultra-strong coupling regime. Under this assumption, states with the same value of oscillator excitation: $\ket{1,0}\ket{N_0}$ and $\ket{1,\pm1}\ket{N_{\pm1}}$, are nearly degenerate. 
We call this quasi-degenerate triplet of states the $N^{th}$ manifold. Because of finite $\omega_0$, there will be transitions between various states: $\ket{1,m}\ket{N_m}$ and $\ket{1,m'}\ket{N^{'}_{m'}}$. These transitions can be classified under two categories: (a) transitions that take place between levels belonging to different manifolds and (b) transitions that take place between the three states that belong to the same manifold. For two adjacent manifolds, transitions of type (a) are shown in Fig. \ref{f.transitions}(a) and for the same two manifolds, transitions of type (b) are shown in Fig. \ref{f.transitions}(b). 
Suppose the states $\ket{1,m}\ket{N_m}$ and $\ket{1,m'}\ket{N^{'}_{m'}}$ belong to different manifolds. The transition matrix element between them is
\beq
\omega_0\left|\matel{1,m}{\hat{S}_{z}}{1,m'}\scpr{N_{m}}{N^{'}_{m'}}\right|.
\eeq
If the above transition matrix element is much smaller than the energy difference between them, i.e. 
\beq
\omega_0\left|\matel{1,m}{\hat{S}_{z}}{1,m'}\scpr{N_{m}}{N^{'}_{m'}}\right|\ll\omega|N-N'|,
\eeq
then the transitions of type (a) would be energetically suppressed. On the other hand, because transitions of type (b) occur between nearly degenerate states, there could be appreciable transfer of population between them. Based on these arguments, one can neglect all matrix elements in $\hat{H}$ that lead to transitions between different manifolds and retain only those terms that induce transitions between states of the same manifold \cite{vanVleck, Shirley-65,Leggett, Irish-05}. This approximation was used by Irish et al. \cite {Irish-05} to study the dynamics of a single quasi-degenerate qubit interacting with a high frequency oscillator.

Under the above assumption of neglecting transitions between states belonging to different manifolds, the Hamiltonian spanned by the $\ket{1,m}\ket{N_m}$ basis reduces to $3\times3$ block diagonal form with each block corresponding to a given manifold. For the $N^{th}$ manifold, the Hamiltonian takes the form 
\beq\label{e.H_N}
 \hat{H}_{N}=
\hbar\omega \begin{pmatrix} N-\beta^2 & \Omega_{N} & 0 \\ 
                \Omega_{N} & N & \Omega_{N}\\
                 0 & \Omega_{N} & N-\beta^2\\
 \end{pmatrix},
\eeq
where rows and columns are arranged in the order: $\ket{1,1}\ket{N_1}$, $\ket{1,0}\ket{N_0}$ and $\ket{1,-1}\ket{N_{-1}}$, and the off-diagonal terms are the normalized Rabi frequencies given by: 
\beqa\label{e.rabi}
\Omega_{N}&=&\frac{\omega_0}{\omega}\matel{1,1}{\hat{S}_{z}}{1,0}\scpr{N_1}{N_0},\nonumber\\
&=&\sqrt{\frac{1}{2}}\frac{\omega_0}{\omega}\la N_{1}|{N_0}\ra, \nonumber \\
&=& \sqrt\frac{1}{2}\frac{\omega_0}{\omega}e^{-\beta^2/2}L_N(\beta^2).
\eeqa
In writing $\hat{H}_{N}$, we have used the fact that $\scpr{N_{1}}{N_0}$ is real and is equal to $\scpr{N_{-1}}{N_0}$. Note that for any value of $N$, $\omega\Omega_N\leq\omega_0$.

We note from the form of $\hat{H}_{N}$ that due to the presence of the off diagonal elements, the state of the oscillator, which is displaced depending upon the state of the qubits, changes with the changing state of the qubits. This change happens in a time scale of $ 1/\left(\omega \Omega_N\right) \geq 1/\omega_0$ which is much slower than the characteristic time scale of the oscillator, which is $1/\omega$. One thus sees that the oscillator state adiabatically adjusts itself to the state of the qubits. For this reason, the above approximation is known as ``adiabatic approximation" \cite{Irish-05,Leggett}. 

The unnormalized eigenfunctions and eigenvalues of $\hat{H}_N$ are:
\beqa\label{e.eigen_omn0}
\ket{\mathcal{E}_{N}^{0}}&=&
\begin{pmatrix} 1 \\ 0 \\ -1\\
\end{pmatrix},\nonumber\\
\ket{\mathcal{E}_{N}^{\pm}}&=&
\begin{pmatrix} 1 \\[.5em] \left(\beta^2\pm\sqrt{8\Omega_{N}^2+\beta^4}\right)/2\Omega_{N} \\[0.5em] 1\\
\end{pmatrix},\nonumber\\
\mathcal{E}_{N}^{0}&=&\hbar\omega(N-\beta^2),\nonumber\\
\mathcal{E}_{N}^{\pm}&=&\frac{\hbar\omega}{2}\left(2N-\beta^2\pm\sqrt{8\Omega_{N}^2+\beta^4}\right).
\eeqa

%===================================================

\section{Population Dynamics}\label{s.collapse_rev}

%===================================================

Analysis of the dynamical properties of a single qubit in the RWA-violating quasi-degenerate regime can be found in \cite{Irish-05, Sandu}. Here we take a step in the direction of $K$-qubit evolution by considering $K$=2, and defer an introduction to cases for $K>2$ to Sec. \ref{s.Nqubits}. We will stay within the quasi-degenerate parameter regime mentioned in Sec. \ref{s.Spectrum} ($\omega_0\leq0.25\omega$) and further make the following assumptions that would considerably simplify the expressions in (\ref{e.eigen_omn0}):
\beq
\Omega_N\gg\beta^2,\ \left |\beta\right |\leq0.2\ {\rm and}\ 
N> 0.
\eeq
This allows some obvious simplifications: $8\Omega_N^2-\beta^4\approx8\Omega_N^2$ and $N-\beta^2\approx N$ respectively. Then the expressions for the eigenfunctions and eigenvalues of $\hat{H}_N$ simplify to:
\beqa\label{e.eigen_omn0_approx}
\ket{\mathcal{E}_{N}^{0}} &=& \frac{1}{\sqrt 2}
\begin{pmatrix} 1 \\ 0 \\ -1\\
\end{pmatrix},\nonumber\\
\ket{\mathcal{E}_{N}^{\pm}} &=& \frac{1}{2}
\begin{pmatrix} 1\\[.5em] \pm\sqrt{2} \\[0.5em] 1\\
\end{pmatrix},\nonumber\\
\mathcal{E}_{N}^{0}&=&\hbar\omega N,\nonumber\\
\mathcal{E}_{N}^{\pm}&=&\hbar\omega\left(N\pm\sqrt{2}\Omega_{N}\right).
\eeqa

For illustration, consider initial states that belongs to the $N^{th}$ manifold:
\beqa
\ket{\Psi_{\pm}(0)} &=& \ket{1,\pm1}\ket{N_{\pm1}}, \nonumber \\
&=& \frac{1}{2}\Big(\ket{\mathcal{E}_{N}^{+}} + \ket{\mathcal{E}_{N}^{-}}\Big) \pm \frac{1}{\sqrt 2}\ket{\mathcal{E}_{N}^{0}}.
\eeqa
Using (\ref{e.eigen_omn0_approx}), the probability amplitude for the qubit to remain in the initial state is easily found to be 
\beq
\la \Psi_{\pm}(0)|\Psi_{\pm}(t)\ra = \frac{e^{-i N\omega t}}{2}\Big(1 + \cos(\sqrt 2 \Omega_N\omega t)\Big).
\eeq
When squared, the probability shows two frequencies of oscillation, $\sqrt 2 \Omega_N\omega$ and $2\sqrt 2 \Omega_N\omega$. Since three basis states are involved, we could expect three frequencies, but two are equal: $|\mathcal{E}_{N}^{+} - \mathcal{E}_{N}^{0}| = |\mathcal{E}_{N}^{-} - \mathcal{E}_{N}^{0}|$. This is in contrast to the single-qubit case where only one Rabi frequency determines the evolution \cite{Irish-05, Sandu}. The two frequencies contribute to the probability as follows:
\beq \label{eqn(1,-1)(t)}
P_{1,_{\pm}1}(N,t) = \frac{3}{8} + \frac{1}{2}\cos(\sqrt2\Omega_N\omega t) + \frac{1}{8}\cos(2\sqrt2\Omega_N\omega t).
\eeq
A different initial state, also characteristic of the two-qubit case, that belongs to the $N^{th}$ manifold is $|1,0\ra|N_0\ra = (\ket{\mathcal{E}_{N}^{+}} + \ket{\mathcal{E}_{N}^{-}})/\sqrt2$. One finds that the probability to remain in this state oscillates with only one frequency, but twice as great as the higher frequency in the previous example:
%, again in distinction to any result for the single qubit:
\beq \label{eqn(1,0)(t)}
P_{1,0}(N,t) = \frac{1}{2} + \frac{1}{2}\cos(4\sqrt2\Omega_N\omega t).
\eeq

A number state is in most cases not a reasonable model for describing experimental excitation of the oscillator. A coherent-state description for the oscillator is more realistic, and in that case the Poisson distribution of number states creates a distribution of Rabi frequencies $\Omega_N\omega$. The oscillations of solutions for different $N$ values rapidly get out of phase with each other and the signal collapses quickly. However, the main contribution of the Poisson distribution comes from its peak near $N = \bar n\approx|\alpha|^2 $ where adjacent Rabi frequencies differ by a small common amount $\delta\Omega(\bar n)\omega$, which leads to a rephasing of the main terms in the summation at integer multiples of the time $2\pi/\delta\Omega(\bar n)\omega$. One thus expects a sequence of revivals and then re-collapses of the signal, which are familiar in parameter regimes where the RWA is valid  \cite{Collapse-Revival}. 

The collapse and revival behavior in the adiabatic approximation for a single qubit case was studied by Irish et al. \cite{Irish-05} and Sandu \cite{Sandu} and we explore here the two-qubit counterpart. We obtain analytical expressions for the collapse and revival times and also for the individual revival signals.   
Since in the two-qubit case the system eigenstates are displaced number states, a coherent-state sum of them produces a displacement of the coherent state $|\alpha\ra$ as the initial state:
\beqa\label{e.initial}
\ket{\Psi_{-}(0)} & = & \ket{1,-1}|\alpha_{-1}\ra \nonumber \\
& \equiv & \ket{1,-1}D(\beta)|\alpha\ra.
\eeqa
Then the probability for the qubits to remain in the state $\ket{1,-1}$ is found to be:
\beq \label{eqnP(t)}
P_{1,-1}(\alpha,t) = \frac{3}{8} + \frac{1}{2}S(t,\omega_0) + \frac{1}{8}S(t,2\omega_0),
\eeq
where 
\beq\label{e.StomDef}
S(t,\omega_0) = \sum_{N=0}^{\infty}\frac{e^{-|\alpha|^2}|\alpha|^{2N}}{N!} \cos{(\omega_0\scpr{N_1}{N_0}t)}.
\eeq
If the average excitation of the oscillator, $|\alpha|^2$, is large, one can evaluate the above sum approximately (see Appendix) to get
\beq\label{e.Stom}
S(t,\omega_0)=Re\left[\sum_{k=0}^{\infty}\bar{S}_{k}(t,\omega_0)\right],
\eeq
where 
\beqa\label{e.dy_S_k}
Re\left[\bar{S}_{k}(t,\omega_0)\right]&=&exp{\left(\frac{-(\tau-\tau_k)^2|\alpha\beta^2|^2}{2\left(1+(\pi kf\right)^2)}\right)}\nonumber\\
&\times&\frac{\cos{(\Phi_{im})}}{\left(1+(\pi kf)^2\right)^{1/4}}.
\eeqa
In (\ref{e.dy_S_k}) we have defined
\beqa
\tau&=&\omega_0 te^{-\beta^2/2},\nonumber\\
f&=&|\alpha\beta|^2,\nonumber\\
\tau_k&=&2\pi k(1+f/2)/\beta^2,
\eeqa
and $\Phi_{Im}$ is given in (\ref{e.A.Phi_im_app}). 

From (\ref{e.Stom}) and (\ref{e.dy_S_k}), it is clear that $S(t,\omega_0)$ exhibits collapse and revival with $\bar S_k(t,\omega_0)$ describing the evolution around the $k^{th}$ revival time.
These individual revival signals, $\bar S_k(t,\omega_0)$, have three salient features: (a) the exponential term in (\ref{e.dy_S_k}) determines the envelope of the revival signal, (b) the cosine term governs the fast oscillatory dynamics and (c) the factor in the denominator determines the height of the $k^{th}$ revival. 
The revival time and the height of the $k^{th}$ revival are:
\beqa\label{e.hight_time}
t_{k}^{rev}=\frac{2\pi k}{\omega_0 \beta^2}(1+|\alpha\beta|^2/2),\nonumber\\
h_{k}=\frac{1}{\left(1+k^2\pi^2|\alpha\beta|^4\right)^{1/4}}.
\eeqa
As usual, the revivals are periodic and the heights of the revivals successively decrease and thus the revivals are never complete. The revival time increases with the increase of the oscillator excitation amplitude, $|\alpha|$, and decreases if the coupling parameter, $|\beta|$, is increased. From (\ref{e.dy_S_k}), we note that the width of the primary revival, for which $k=0$, is 
\beq
\delta\tau_{0}=\frac{1}{|\alpha| \beta^2},
\eeq
and the width of the $k^{th}$ revival is given by
\beq\label{e.width}
\delta\tau_k=\delta\tau_0\sqrt{1+(\pi k|\alpha\beta|^2)^2}.
\eeq
Thus, we see that the width of the successive revival signals keep increasing. We note that the term $|\alpha\beta|^2/2$ in the expression for the revival time (\ref{e.hight_time}), is an improvement over the results given by Irish et al. \cite{Irish-05} and Sandu \cite{Sandu}.

\begin{figure}[htb]
\includegraphics[width=8cm,height=3cm]{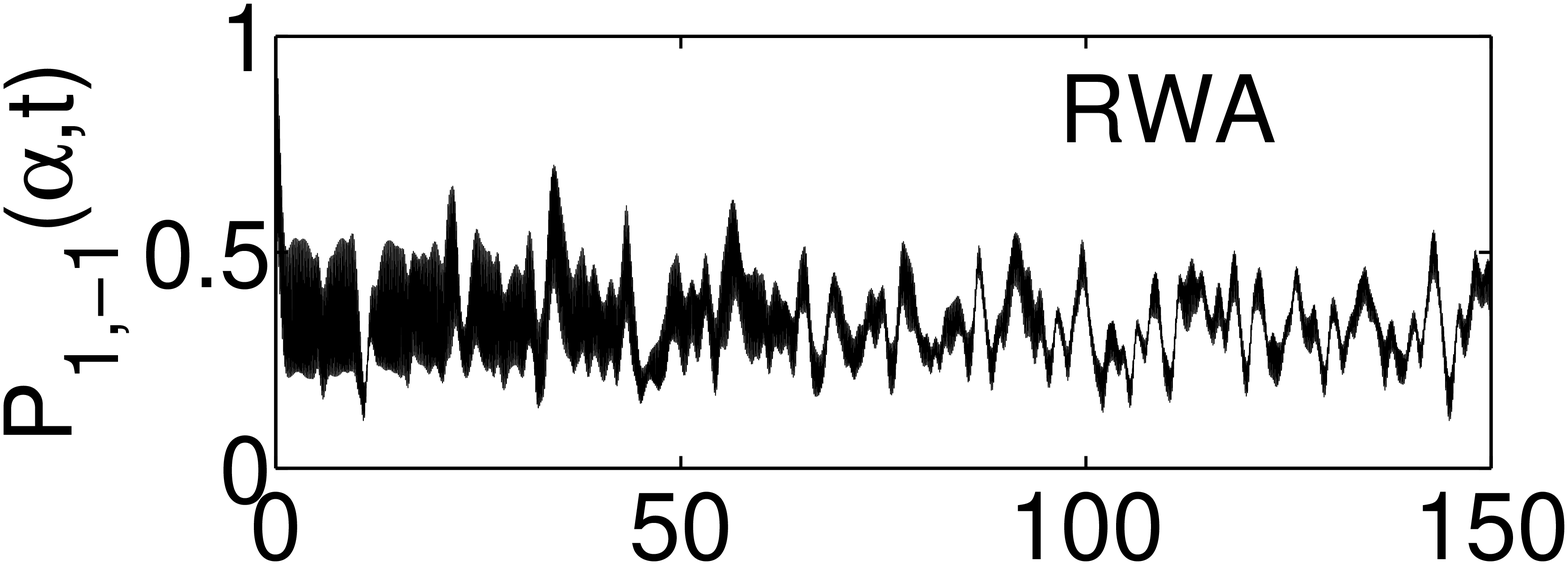}

\vspace{-0.5cm}
\includegraphics[width=8cm,height=3cm]{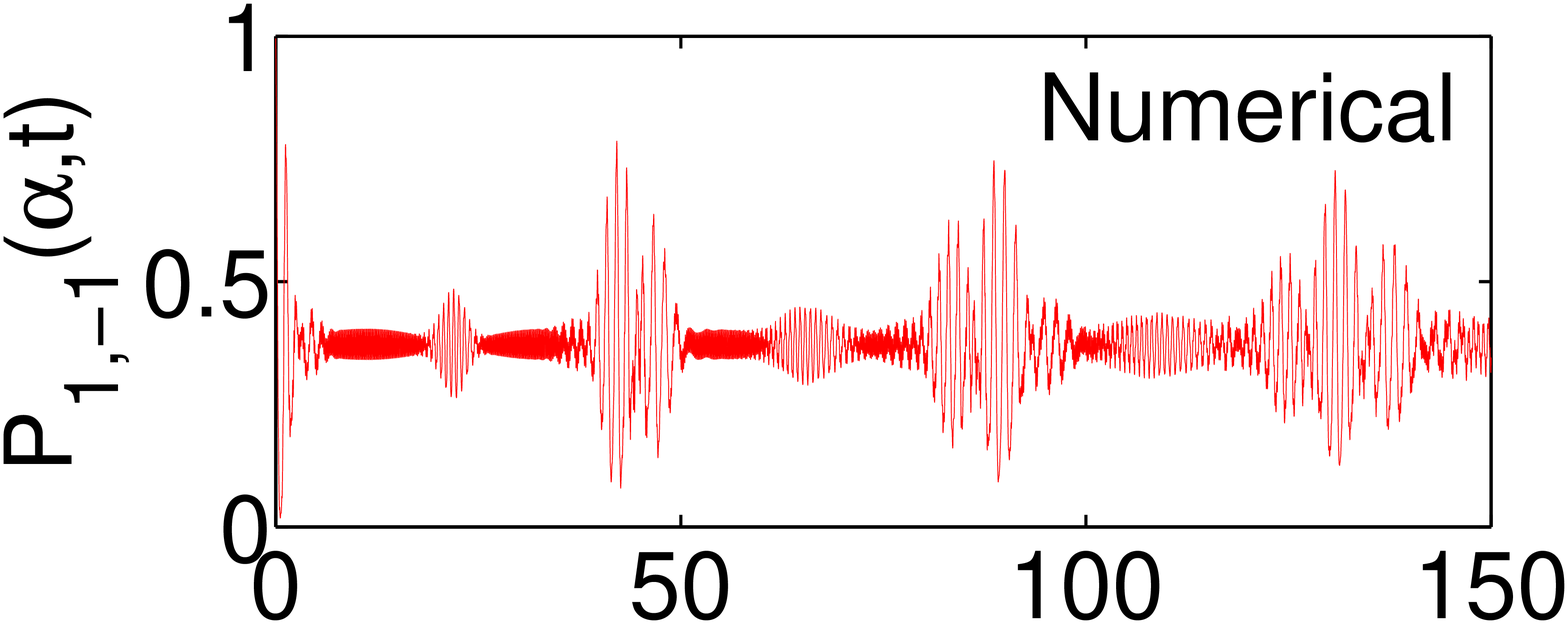}

\vspace{-0.5cm}
\includegraphics[width=8cm,height=3cm]{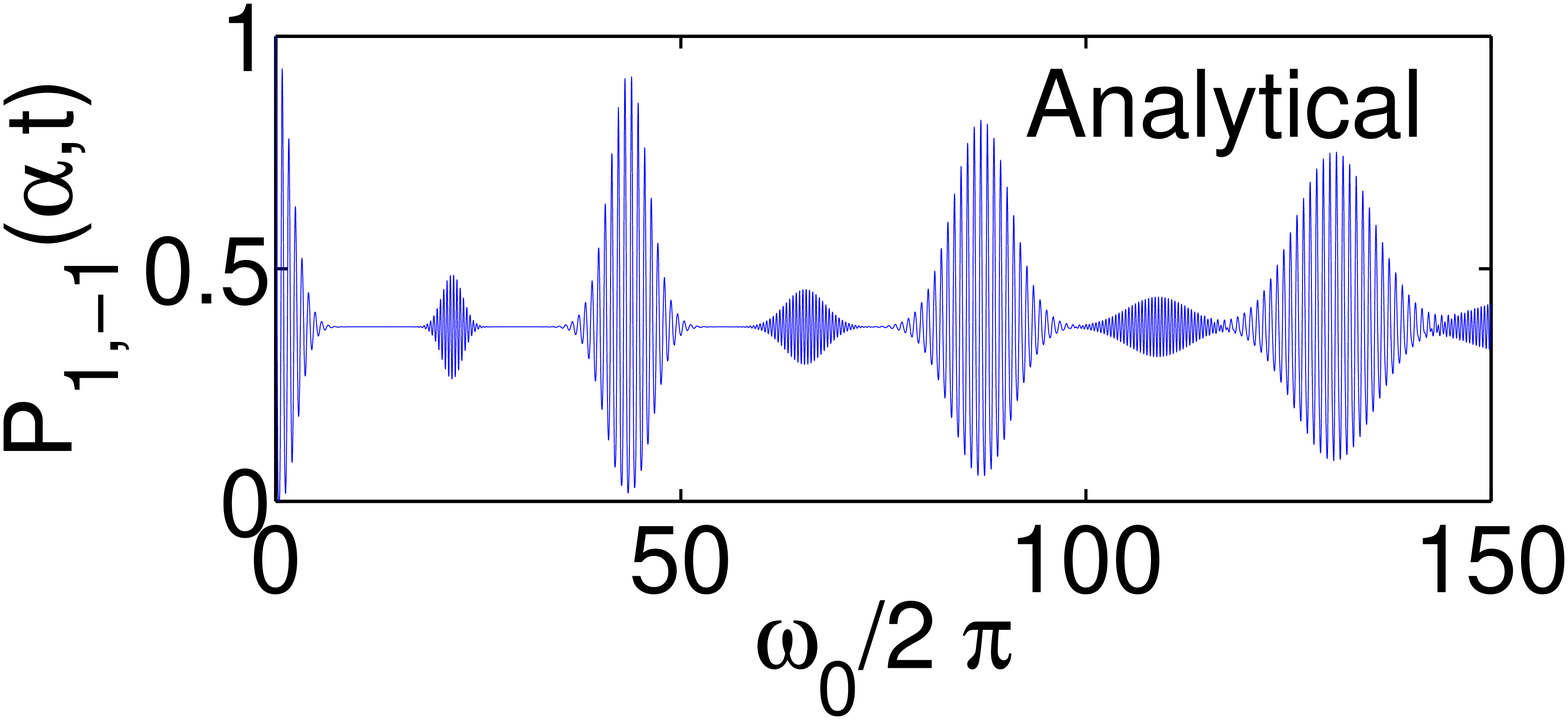}
\caption{Collapse and revival dynamics for $P_{1,-1}(\alpha,t)$, given $\omega_0 = 0.15\omega$, $\beta=0.16$ and $\alpha=3$. Note the breakup in the main revival peaks of the numerical evaluation, which comes from the $\omega_0$-$2\omega_0$ beat note, not included in the analytic calculation. The RWA is seen to break down completely in the parameter regime considered.}
\label{f.collapse_revival_double} 
\end{figure}

From (\ref{eqnP(t)}) we note that two functions are responsible for the evolution of $P_{1,-1}(\alpha,t)$: $S(t,\omega_0)$ and $S(t,2\omega_0)$. Thus, we get two different revival sequences in the evolution of $P_{1,-1}(\alpha,t)$. The analytic formula derived for $P_{1,-1}(\alpha,t)$ is plotted in Fig. \ref{f.collapse_revival_double} and is compared with the numerical calculations.  
The revival signals corresponding to the terms $\bar{S}_{k}(t,\omega_0)$ and $\bar{S}_{2k}(t,2\omega_0)$ overlap in time and produce a beat note. 
This is evident in Fig. \ref{f.collapse_revival_double}. As mentioned earlier, the RWA completely breaks down in the parameter regime we consider. This can clearly be seen in Fig. \ref{f.collapse_revival_double} where the evolution of $P_{1,-1}(\alpha,t)$ calculated within the RWA disagrees even qualitatively with the numerical calculations.  

The expression (\ref{e.dy_S_k}), was derived under the constraint $|\alpha\beta|\ll1$ (see (\ref{e.A.app_Om})). Within this constraint, the revival time was found to be a monotonically increasing and decreasing function of $|\alpha|$ and $|\beta|$ respectively. If the oscillator excitation number and the coupling strength are not restricted by the constraint $|\alpha\beta|\ll1$, the revival time is no longer a monotonic function of $|\alpha|$ and $|\beta|$. This non-monotonic behavior was numerically explored in \cite{Irish-05}. In the limit of very high oscillator excitation number, we can employ the asymptotic expression for $L_{\bar{n}}(\beta^2)$ to derive an analytic expression for the revival time of $S(t,\omega_0)$ that is not restricted by the constraint $|\alpha\beta|\ll1$.

As we already mentioned, revivals should occur at multiples of the time $t=t^{rev}$ such that
\beqa\label{e.Rev}
&&\delta\Omega(\bar n)\omega t^{rev}=2\pi, \ {\rm or} \nonumber\\
&&\omega_0e^{-\beta^2/2}\left|L_{\bar{n}+1}(\beta^2)-L_{\bar{n}}(\beta^2)\right|t^{rev}=2\pi.
\eeqa
If the oscillator is highly excited, $\bar{n}\gg1$, 
one can use the following asymptotic formula for the Laguerre polynomial \cite{Abramowitz}:
\beq\label{e.asymp_Laguerre}
\lim_{\bar{n}\to\infty}e^{-x/2}L_{\bar{n}}(x)=\frac{\cos{(2\sqrt{\bar{n}x}-\pi/4)}}{\sqrt{\pi}(\bar{n}x)^{1/4}},
\eeq
to obtain:
\begin{align}\label{e.rev_time}
\left(\frac{\omega_0 t^{rev}}{2\pi}\right)^{-1}=&\Big|\frac{\cos{(2|\alpha\beta|-\pi/4)}}{\sqrt{\pi |\alpha^5\beta|}}\nonumber\\
&+\sqrt{\frac{|\beta|}{\pi|\alpha^3|}}\sin{(2|\alpha\beta|-\pi/4)}\Big|.
\end{align} 
From (\ref{e.rev_time}), the non-monotonic dependence on $\alpha$ and $\beta$ of the revival time is clear. Note that equation (\ref{e.rev_time}) is not restricted by the constraint $|\alpha\beta|\ll1$. 

\begin{figure}[htb]
	\centering
	\includegraphics[width=8cm,height=2.5cm]{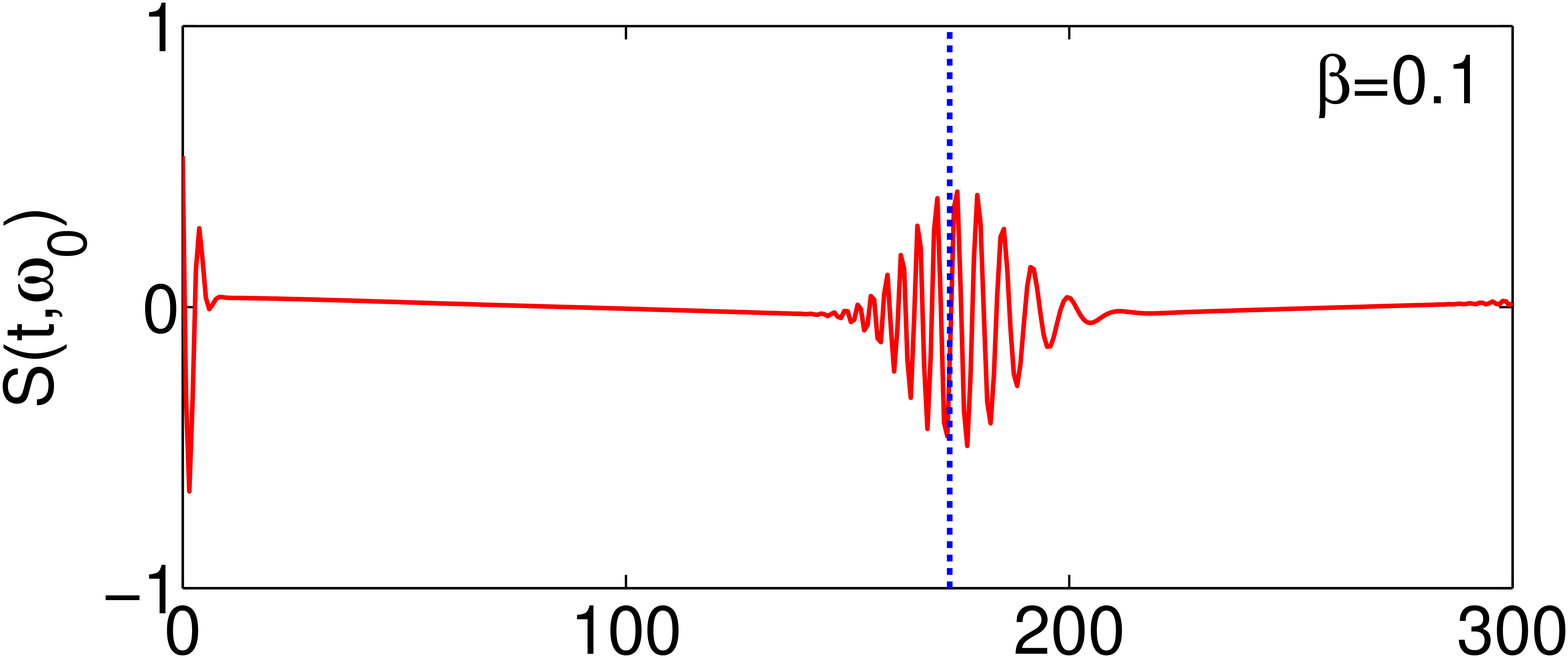}
	\includegraphics[width=8cm,height=2.5cm]{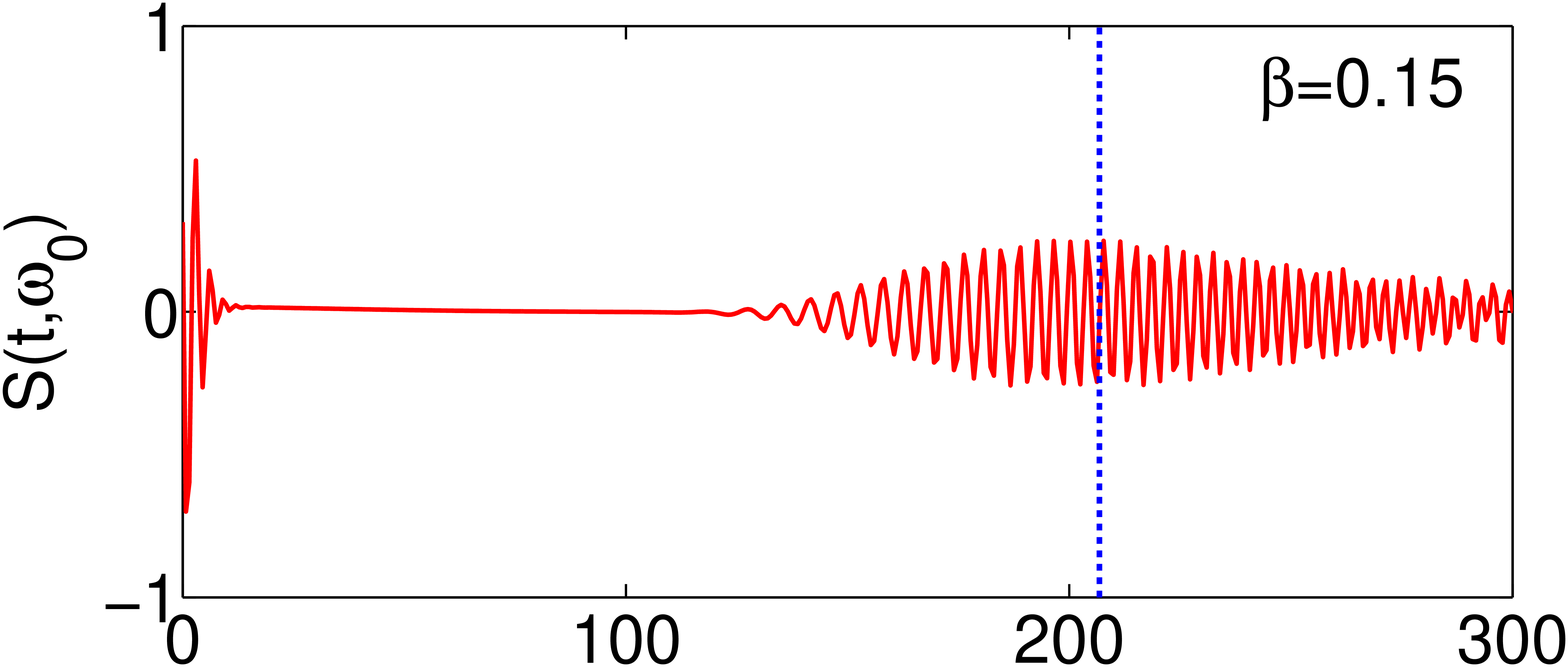}
	\includegraphics[width=8cm,height=2.5cm]{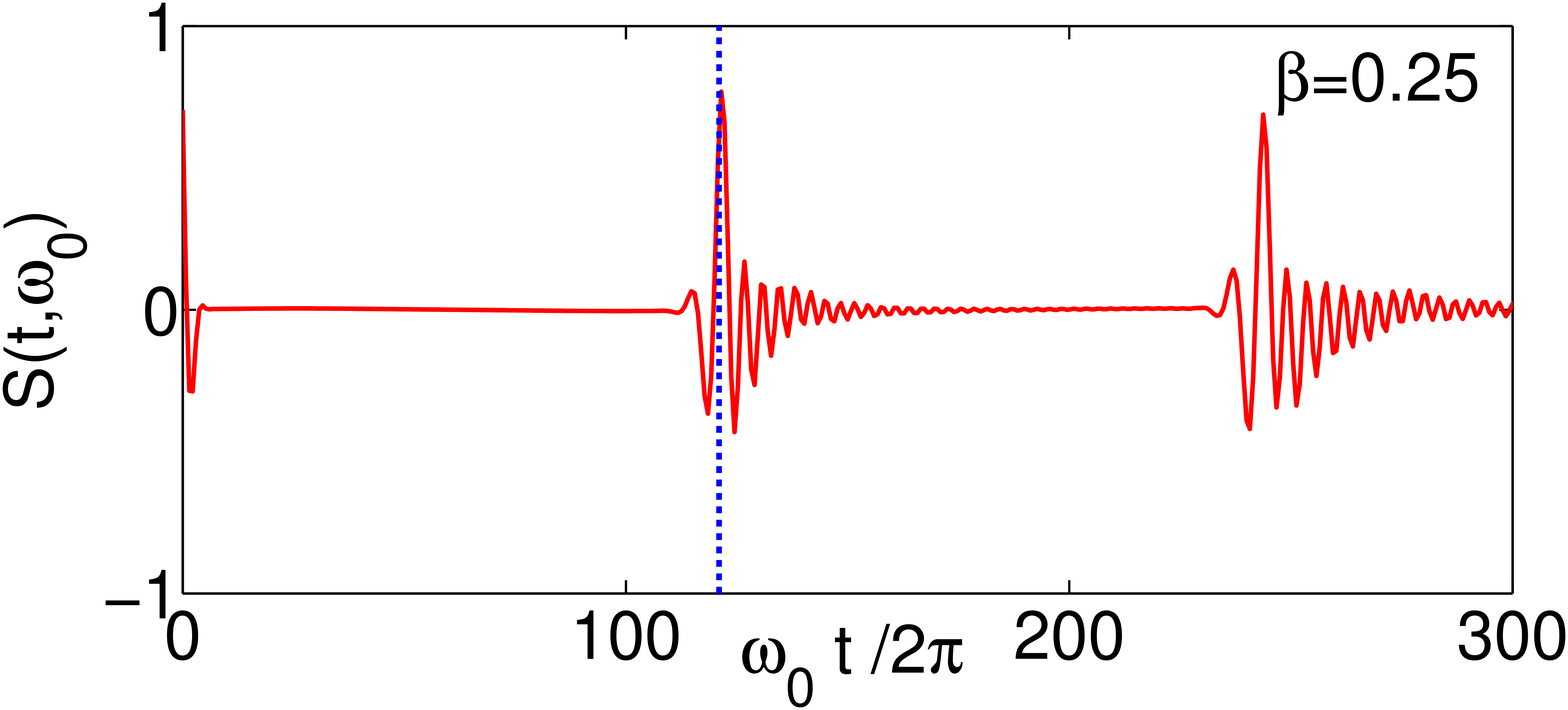}
	\caption{Numerical evaluation of $S(t,\omega_0)$ for $\alpha=10$ and various values of $\beta$. Note that the revival time is a non-monoitonic function of $\beta$. The vertical lines correspond to the revival time predicted by Eq. (\ref{e.rev_time}) and are seen to coincide with the numerically evaluated revival signals.}
	\label{f.col_rev_big_alpha}
\end{figure}

In Fig. \ref{f.col_rev_big_alpha}, we plot $S(t,\omega_0)$ for $\alpha=10$ and various values of $\beta$. The revival times predicted by (\ref{e.rev_time}) are denoted by vertical lines and are seen to have excellent agreement with the numerically evaluated revival signals despite a strongly varying location of the revivals. 
The figure clearly demonstrates the non-monotonic dependence of the revival time on the coupling strength and also highlights the departure from the formula for $t^{rev}$ derived in Sec. \ref{s.collapse_rev} under the constraint $|\alpha\beta|\ll1$. The revival envelopes seen in Fig. \ref{f.col_rev_big_alpha} are not approximately gaussians as was the case for the revivals studied in Sec. \ref{s.collapse_rev}. A detailed discussion of the non-trivial structure of the revivals for big values of $|\alpha|$ and $|\beta|$ can be found in \cite{Irish-05}.

We now contrast the dynamical evolution of the two-qubit TC model with the single qubit system. We assume that the initial state of the single qubit system is:
\beq
\ket{\Psi_{s}(0)}=\ket{1/2,-1/2}\ket{\alpha_{-{1/2}}},
\eeq
where $\hat\sigma_x\ket{1/2,-1/2}=-\ket{1/2,-1/2}$ and $\hat D(\beta/2)\ket{\alpha}=\ket{\alpha_{-1/2}}$.
The state $\ket{\Psi_{s}(0)}$ is a single qubit counterpart of the two-qubit initial state, $\ket{\Psi_{-}(0)}$ given in (\ref{e.initial}), in the sense that both the states are such that (a) the qubit(s) and the oscillator are initially uncorrelated, (b) the qubit(s) state is an eigenstate of $\hat{S}_{x}$ with the lowest possible eigenvalue of $m$ and (c) the oscillator is in a displaced coherent state. Under the adiabatic approximation, the evolution of the state can be analytically derived  \cite{Irish-05} and one finds that the probability for the qubit to be in the state $\ket{1/2,-1/2}$ evolves as:
\begin{align}\label{e.Pmm}
P_{\frac{1}{2},-\frac{1}{2}}(\alpha,t)&=1/2\left(1+\sum_{N=0}^{\infty}e^{-|\alpha|^2}\frac{|\alpha|^{2N}}{N!}\cos{\left(\sqrt{2}\omega\Omega_{N}t\right)}\right),\nonumber\\
&=1/2\Big(1+S(t,\omega_0)\Big).
\end{align}
\begin{figure}[htb]
\includegraphics[width=8cm,height=3cm]{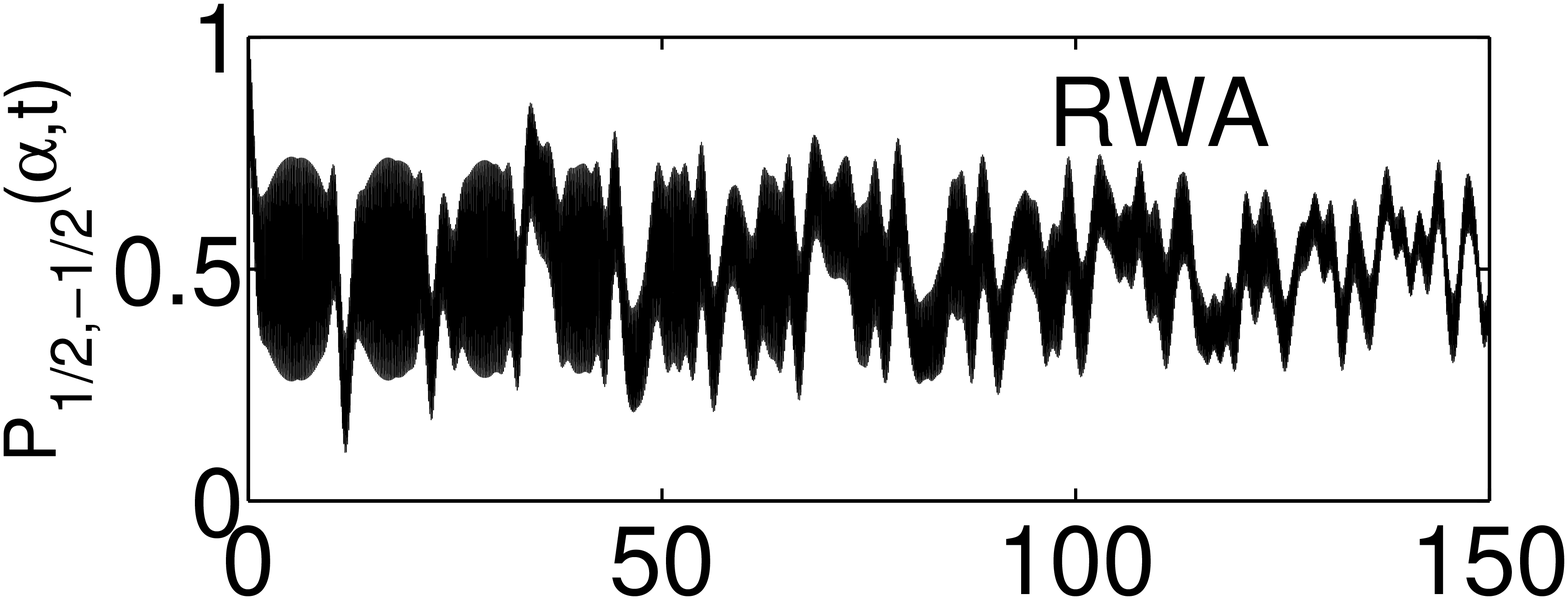}

\vspace{-0.5cm}
\includegraphics[width=8cm,height=3cm]{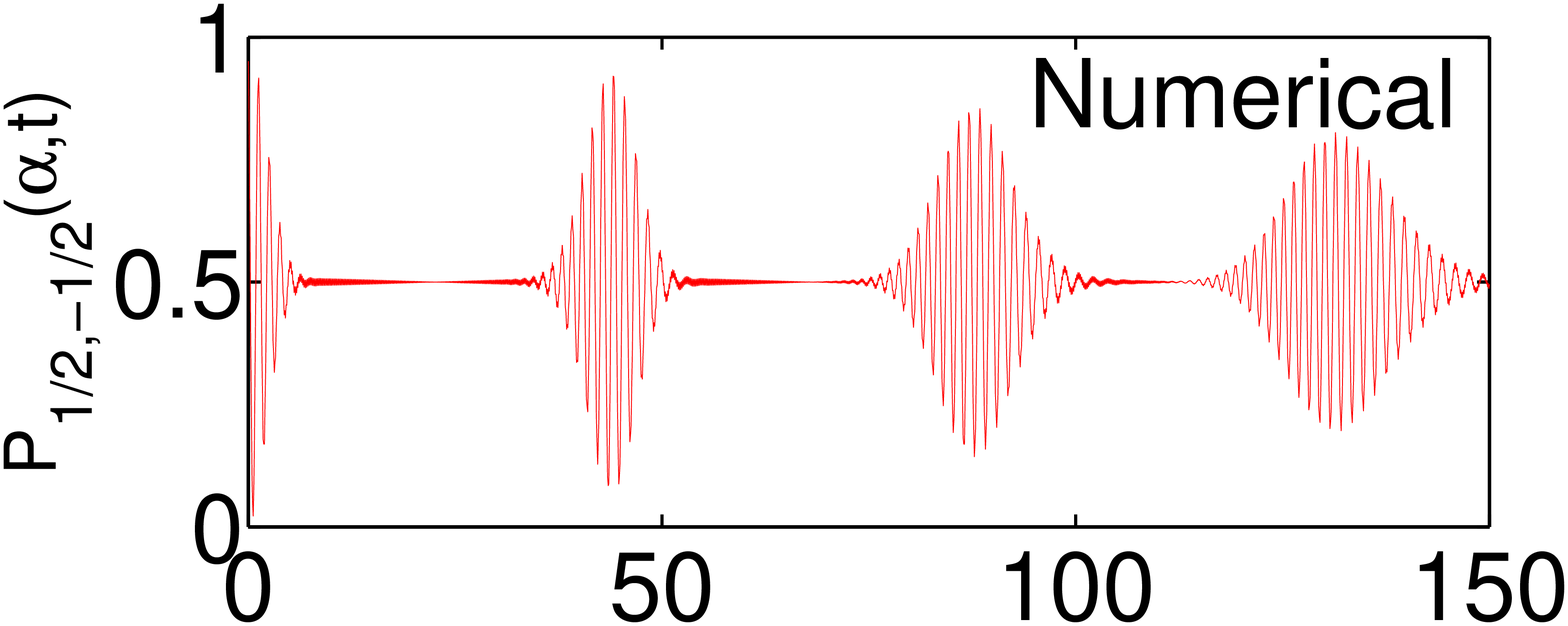}

\vspace{-0.2cm}
\includegraphics[width=8cm,height=3cm]{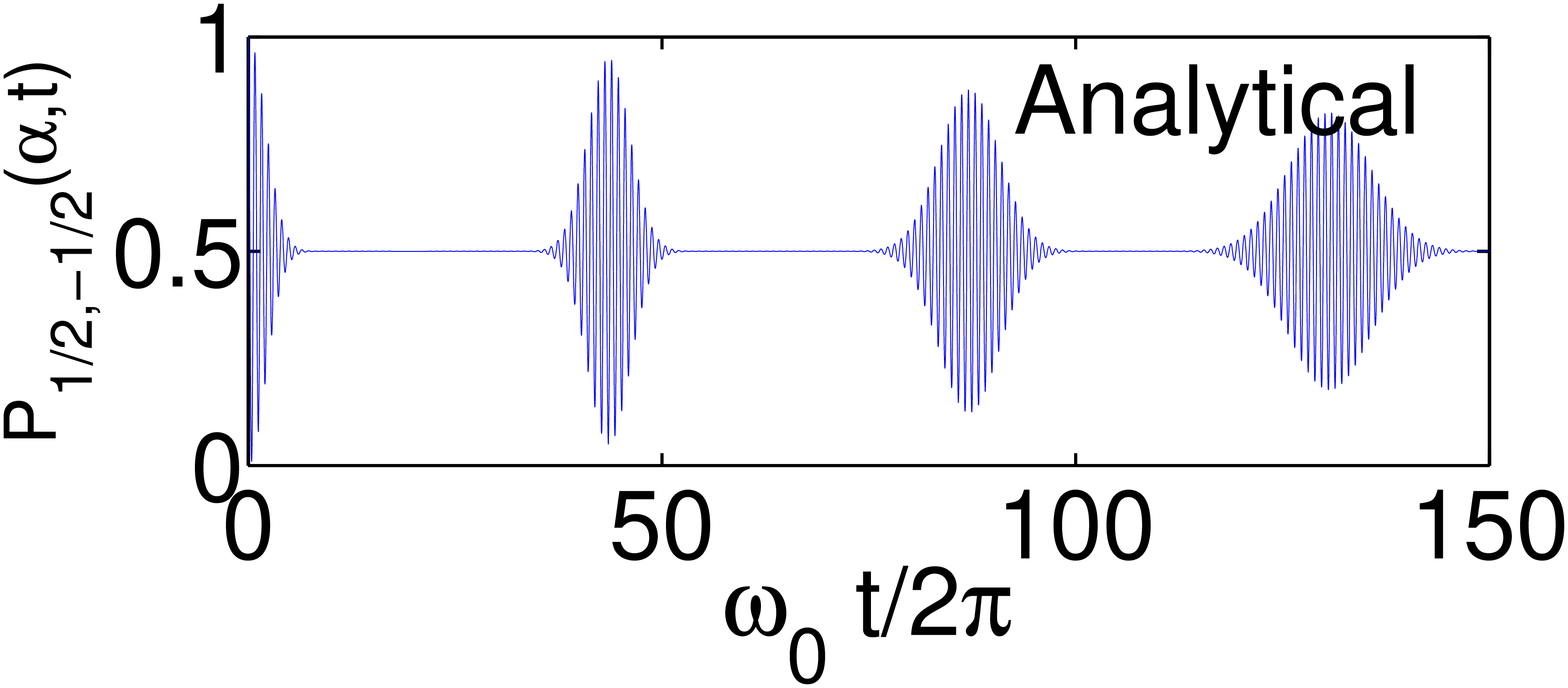}
\caption{Collapse and revival dynamics for $P_{\frac{1}{2},-\frac{1}{2}}(\alpha,t)$, given $\omega_0 = 0.15\omega$, $\beta=0.16$ and $\alpha=3$. Note the single revival sequence. Also, note that there are no breakups in the revival peaks in contrast to the two-qubit case (Fig. \ref{f.collapse_revival_double}). The RWA fails to describe the dynamical evolution even for the single qubit case.}
\label{f.collapse_revival_single} 
\end{figure}
There is only one revival sequence for the single qubit system as a consequence of having only one Rabi frequency in the single qubit case.
The analytic and numerically exact evolution of $P_{\frac{1}{2},-\frac{1}{2}}(\alpha,t)$ is plotted in Fig. \ref{f.collapse_revival_single}. The single revival sequence is evident from the figure. A discussion on the multiple revival sequences for the $K$-qubit TC model, within the parameter regime where the RWA is valid, can be found in \cite{Deng}. 

%===================================================

%\section{Entanglement Dynamics}\label{s.entanglement}

%===================================================

\section{Entanglement Dynamics}\label{s.entanglement}
The evolution of entanglement between several non-interacting qubits coupled to a single mode or many oscillator modes, which act like quantum buses mediating information between the qubits, has been studied extensively (e.g., see \cite{Lee-06,Yonac-10, Tessier}). In all these cases, the interaction between the qubits and the oscillator mode(s) were treated within the RWA. New time scales and qualitatively new features arise in regimes where the RWA is invalid, mandating an extension of these previous results \cite{Jing-Ficek, Ficek_etal, Chen_etal, Leon-Sabin}. Applications may be important in areas of quantum information processing, where coherent control of entanglement may be vital.

A generic illustration can start in a configuration without correlation between the oscillator and qubits. We place the qubits in one of the $\hat\sigma_x$ Bell states and the oscillator in an undisplaced coherent state $|\alpha\ra$:
\beqa\label{e.ini_state}
\ket{\xi(0)}&=&\frac{1}{\sqrt{2}}\left(\ket{++}+\ket{--}\right)\ket{\alpha},\nonumber\\
&=&\frac{1}{\sqrt{2}}\left(\ket{1,1}+\ket{1,-1}\right)\ket{\alpha}.
\eeqa
We continue with the approximations used to study the evolution of $P_{1,-1}(\alpha,t)$ in the previous section. In particular, we focus on the changes arising from the presence of four effective oscillators. Then, given $N \gg \beta$, we make the following approximation:
\beq\label{e.smallbeta}
\ket{N_0}\approx\hat{D}(\mp\beta)\ket{N_0}=\ket{N_{\pm1}},
\eeq
which leads to
\beqa
\ket{\xi(0)} = \frac{1}{\sqrt{2}}\left(\ket{1,1}\ket{\alpha_{1}} +\ket{1,-1}\ket{\alpha_{-1}}\right).
\eeqa
Using (\ref{e.eigen_omn0_approx}), one evaluates the time evolved state to be:
\beqa
\ket{\xi(t)}=&&\frac{e^{-|\alpha|^2/2}}{2\sqrt{2}}\sum_{N=0}^{\infty}\frac{\alpha^N}{\sqrt{N!}}\Big[(e^{-i\mathcal{E}_N^{+}t/\hbar}+e^{-i\mathcal{E}_N^{-}t/\hbar})\nonumber\\
&&\times\left(\ket{1,1}\ket{N_1}+\ket{1,-1}\ket{N_{-1}}\right)\nonumber\\
&&+\sqrt{2}(e^{-i\mathcal{E}_N^{+}t/\hbar}-e^{-i\mathcal{E}_N^{-}t/\hbar})\ket{1,0}\ket{N_0}\Big],\nonumber
\eeqa
and from (\ref{e.smallbeta}) one sees that this reduces to:

\beqa
\ket{\xi(t)} &=& \frac{e^{-|\alpha|^2/2}}{2\sqrt{2}}\sum_{N=0}^{\infty}\frac{\alpha^N}{\sqrt{N!}} \nonumber \\
&\times&\Big[(e^{-i\mathcal{E}_N^{+}t/\hbar} + e^{-i\mathcal{E}_N^{-}t/\hbar}) \left(\ket{1,1} + \ket{1,-1}\right)\nonumber\\
&+& \sqrt{2}(e^{-i\mathcal{E}_N^{+}t/\hbar} - e^{-i\mathcal{E}_N^{-}t/\hbar}) \ket{1,0}\Big]\ket{N_0}.\nonumber
\eeqa
This is particularly compact in the $\hat{\sigma}_{z}$ eigenbasis ($\hat{\sigma}_{z}\ket{e}=\ket{e}$, $\hat{\sigma}_{z}\ket{g}=-\ket{g}$), where it becomes:
\beqa \label{Phi(t)2}
\ket{\xi(t)} &=& \frac{e^{-|\alpha|^2/2}}{\sqrt{2}} \sum_{N=0}^{\infty}\frac{\alpha^N}{\sqrt{N!}} \nonumber\\
&\times& \left(e^{-i\mathcal{E}_N^{+}t/\hbar}\ket{ee} 
+ e^{-i\mathcal{E}_N^{-}t/\hbar}\ket{gg}\right)\ket{N_0}.
\eeqa

\begin{figure}[htb]
	\centering
	\includegraphics[width=9cm]{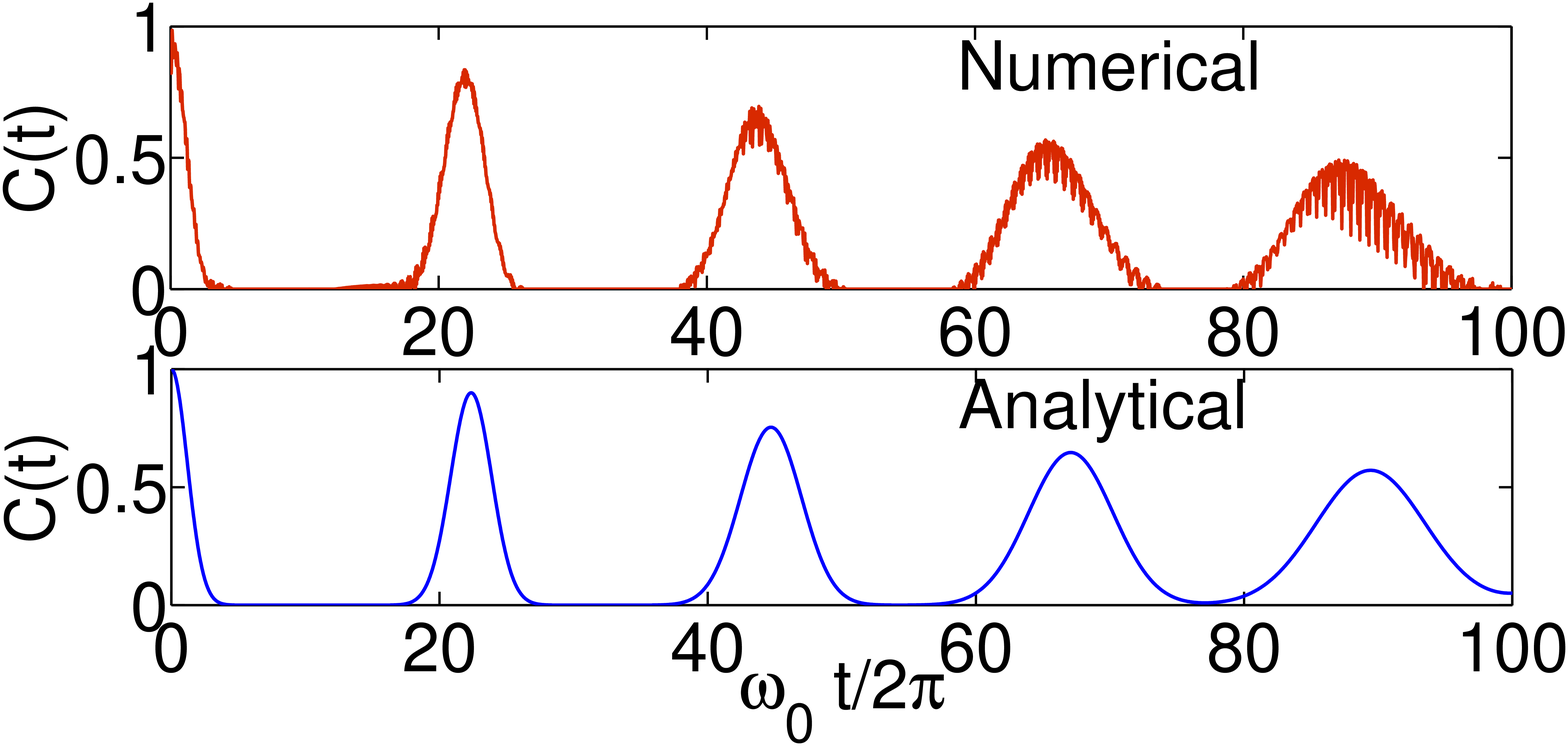}
	\caption{Numerical and analytical evaluation of the entanglement dynamics between the two qubits for $\omega_0=0.15\omega$, $\beta=0.16$ and $\alpha=3$. Entanglement between the qubits exhibits collapse and revival. The analytic expression agrees well with the envelope of the numerically evaluated entanglement evolution.}
	\label{f.ent_col_rev}
\end{figure}

In order to study the entanglement dynamics between the two qubits, we first trace out the oscillator degrees of freedom to get the reduced density matrix for the qubits: 
\begin{align}\label{e.rho_12}
 \hat{\rho}^{(1,2)}(t)&=\sum_{N_0}\scpr{N_0}{\xi(t)}\scpr{\xi(t)}{N_0},\nonumber\\
 &=\frac{1}{2} \Big(\ket{ee}\bra{ee}+\ket{gg}\bra{gg}\Big)\nonumber\\
 &\quad+\frac{1}{2}\Big(\sum_{k=0}^{\infty}\bar{S}_{k}(t,2\omega_0)\ket{gg}\bra{ee}+\mathrm{H.c.}\Big).
\end{align}
At time $t=0$, the state of the qubits is pure, but as time evolves, the reduced state of the qubits becomes mixed. One can use concurrence to quantify entanglement between two qubits that are in an arbitrary mixed state \cite{Wootters}. Concurrence varies in the range from zero to one with zero denoting no entanglement and one denoting maximum entanglement between the qubits. The density matrix, $\hat{\rho}^{(1,2)}(t)$, is an example of a so called X-matrix \cite{X-Matrix}. Calculating the concurrence for an X-matrix is particularly easy and for $\hat{\rho}^{(1,2)}(t)$ it is evaluated to be:
\beq
C(t)=\left|\sum_{k=0}^{\infty}\bar{S}_{k}(t,2\omega_0)\right|.
\eeq 
As discussed in Sec. \ref{s.collapse_rev}, this expression has periodic revivals with each term, $\bar{S}_{k}(t,2\omega_0)$, in the sum centered at the $k^{th}$ revival. When they are well resolved and don't overlap and one is only interested in the envelope of the revivals, one can neglect the interference between the various terms in the sum and an approximate expression for the concurrence is found to be:
\begin{align}\label{e.con}
C(t)&\approx\sum_{k=0}^{\infty}\left|\bar{S}_{k}(t,2\omega_0)\right|,\nonumber\\
&=\sum_{k=0}^{\infty}\frac{1}{\left(1+(\pi kf)^2\right)^{1/4}}\nonumber\\
&\times exp{\left(\frac{-(2\tau-\tau_k)^2|\alpha\beta^2|^2}{2\left(1+(\pi kf\right)^2)}\right)}.
\end{align}
This expression for concurrence is plotted in Fig. \ref{f.ent_col_rev}. We see that the entanglement between the qubits exhibits collapse and revival and the analytic formula agrees well with the envelope of the numerically evaluated result, predicting correctly the time, height and width of the individual entanglement revival signals. 

%===================================================

%\section{Beyond the two qubits case}\label{s.Nqubits}

%===================================================

\section{Generalization to $K$-qubit system}\label{s.Nqubits}
The analysis presented so far is restricted to the two qubits case. In this section, we will qualitatively sketch the procedure for extending the formalism of studying two quasi-degenerate qubits interacting with a high frequency oscillator to the $K$-qubit system. The Hamiltonian governing the dynamics of the $K$-qubit TC model is an obvious generalization of (\ref{RabiHam}), where $\hat{\sigma}$ operators are replaced by their $K$-qubit counterparts:
\beq\label{NHamiltonian}
\hat{H}=\hbar\omega_{0}\hat{S}_{z} + \hbar\omega \hat{a}^{\dagger}\hat{a}
+ \hbar\omega\beta(\hat{a}+\hat{a}^{\dagger})\hat{S}_{x},
\eeq
where
\beq
\hat{S}_{z} = \frac{1}{2}\sum_{i=1}^{K}\hat{\sigma}_{z}^{(i)},\ {\rm and}\ 
\hat{S}_{x} = \frac{1}{2}\sum_{i=1}^{K}\hat{\sigma}_{x}^{(i)}.
\eeq
The eigenfunctions of $\hat{H}$ when $\omega_0=0$ are $\ket{j,m}\ket{\Phi_m}$, where the states $\ket{j,m}$ are eigenfunctions of $\hat{S}_{x}$ and the states $\ket{\Phi_m}\equiv\ket{N_m}$ are the generalization of the displaced Fock states defined in (\ref{e.dis_Fock}). For simplicity in notation we will take $K$ to be even, and then we have
\beq 
j = 0,1,\dots, K/2 ,\ {\rm with}\ m = -j, \dots, j.
\eeq
The eigenvalue of $\hat H$ corresponding to the state $\ket{j,m}\ket{N_m}$ is the same as before, $E_{N,m} = \hbar\omega (N-m^2\beta^2)$, which is independent of $K$. If we assume that $(K/2)^2\beta^2\ll 1$, the states with the same value of oscillator excitation number, $N$, will have nearly the same energy and will form a quasi-degenerate manifold. For a given $j$, a quasi-degenerate manifold consists of 
$2j+1$ states (one state for each $m$). 

The matrix elements 
of the perturbation term, $\omega_0 \hat{S}_{z}$,  in the $\ket{j,m}\ket{N_m}$ basis are:
\begin{align}\label{e.pertub_mat_el}
\omega_0&\matel{j,m}{\hat{S}_{z}}{j',m'}\scpr{N_m}{N'_{m'}}=\omega_0\frac{\delta_{j,j'}}{2}\scpr{N_m}{N'_{m'}}\nonumber\\
\times&\left(\delta_{m'+1,m}\sqrt{(j-m)(j+m+1)}\right.\nonumber\\
&\quad\left.+\delta_{m'-1,m}\sqrt{(j+m)(j-m+1)}\right).
\end{align}
We see from the above formula that the perturbation Hamiltonian connects only the states with the same value of total spin $j$. 
Under the adiabatic approximation, we retain only those terms of the perturbation Hamiltonian for which $N=N'$. Thus the Hamiltonian for the $K$-qubit case breaks into $(K/2)+1$ disjoint Hamiltonians with each disjoint Hamiltonian corresponding to a given $j$. 
This is a consequence of having $\delta_{j,j'}$ in equation (\ref{e.pertub_mat_el}).
Furthermore, for a given $j$, the Hamiltonian assumes a block diagonal form in the $\ket{j,m}\ket{N_m}$ basis, with each block corresponding to a particular value of $N$. Each of these blocks has a dimension of $(2j+1)\times(2j+1)$. This block diagonalization of the Hamiltonian is a consequence of the adiabatic approximation. 
Finding the eigenvalues and eigenfunctions of each of these block diagonal matrices allows one to study the dynamical evolution of the system analytically.

Entanglement dynamics of the $K$-qubit TC model is important in understanding multi-particle quantum coherences. 
Multipartite entanglement is still largely mysterious, in the sense 
that no approach is known that provides both necessary and sufficient 
criteria for arbitrary mixed-state entanglement of more than two parties or 
even for just two parties if their Hilbert states have dimensions 
greater than 2 $\times$ 3 \cite{Horodecki}. 

The pure-state situation is clearer. In that case one can use the so called Schmidt weight to reliably quantify entanglement between two-party states of arbitrary dimensions \cite{Grobe}. 
Thus, if our initial state is pure, we can conceptually partition the composite system into two parties and study the entanglement dynamics between them by using Schmidt analysis. 
Note that our composite system consists of $K+1$ parties ($K$ qubits and an oscillator) and so there can be $2^K-1$ different bi-partitions.

%===================================================

%\section{Validity Region}\label{s.validity}

%===================================================

\section{Validity Region}\label{s.validity}

\begin{figure}[htb]
	\centering
	\includegraphics[width=8.5cm]{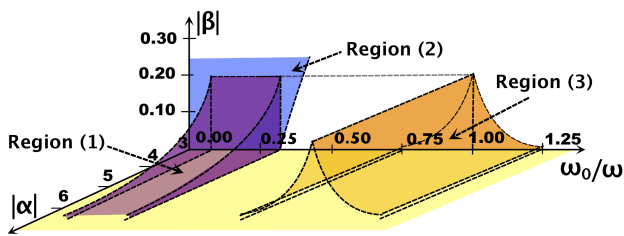}
	\caption{Region (1): Parameter regime where the analytic formula for the collapse and revival dynamics derived within the adiabatic approximation is valid. Region (2): Parameter regime where the eigenspectrum derived within the adiabatic approximation is valid. Region (3): Parameter regime where the RWA is valid.}
	\label{f.validity}
\end{figure}

For better appreciation of the zones of validity of the different approximate approaches to the RWA and quasi-degenerate evolution dynamics, we show a 3D representation in Fig. \ref{f.validity}. The three axes in the figure correspond to the three key dimensionless parameters: $|\beta|$, $\omega_0/\omega$ and $|\alpha|$. Region (1) is the zone where the formula for the individual revival signals (\ref{e.dy_S_k}) derived within the adiabatic approximation is valid, and to summarize, we list here the restrictions on validity:
\begin{itemize}
\item{$\omega_0\leq0.25\omega$: The adiabatic approximation is valid.}
\item{$\Omega_{\bar{n}}\gg\beta^2$: Necessary for the eigenvalue and eigenfunction simplifications leading to (\ref{e.eigen_omn0_approx}).}
\item{$|\alpha|\gg 1$: Validates the assumptions made in the Appendix concerning $\bar{S}_{k}(t)$.}
\item{$|\beta|\leq0.2$: Imposed according to current experimental realizability. It is necessary for nearly degenerate states to have the same value of oscillator excitation number, $N$, and also for the simplifications that lead to (\ref{e.eigen_omn0_approx}).}
\item{$|\alpha\beta|\ll1$: Necessary for restricting the power series expansion of the Laguerre polynomial, $L_N$, to the first three terms (see (\ref{e.A.app_Om}))}.
\end{itemize}
For $|\beta|\leq0.25$, region (2) in Fig. \ref{f.validity} corresponds to the parameter regime where the eigen-spectrum of the system derived within the adiabatic approximation is valid. 
In region (3) of Fig. \ref{f.validity} we show the regime where the analytic formula for the collapse and revival signals of two-qubit TC model derived within the RWA \cite{Deng} is valid. At resonance, the RWA results are valid for coupling strength as big as $|\beta|=0.2$. With the increase of detuning, the validity of the RWA is restricted to lower values of the coupling strength. We see that regions (1) and (3) are completely disjoint and thus the dynamics predicted within the adiabatic approximation cannot be derived from RWA calculations.

%===================================================

%\section{Conclusion}\label{s.conclusion}

%===================================================

\section{Conclusion}\label{s.conclusion}

In this report we extended the Tavis-Cummings model for multi-qubit interaction with a common oscillator beyond its familiar RWA limits, in order to acomplish two goals: (a) to analyze two-qubit dynamics in the quasi-degenerate regime, following seminal work for a single qubit and oscillator by Irish \cite{Irish-05, Irish-07} others \cite{Hausinger-10, Ashhab, Sandu}, and (b) to extend studies in this domain to include the dynamical behavior of quantum coherence in the form of qubit-qubit entanglement. We restricted attention to the regime where the qubit frequencies are much smaller than the oscillator frequency and the interaction coupling energy between the qubits and the oscillator is allowed to be an appreciable fraction of an oscillator energy quantum. 

We worked within the same adiabatic approximation introduced by Irish et al. \cite{Irish-05}. We showed that the RWA and quasi-degenerate regions do not overlap, and that the expressions derived lie completely outside the validity regime of the RWA. We were able to compare features of single qubit dynamics with the corresponding extended results for two qubits, to identify features that are a consequence of having multiple qubits in the system and are qualitatively different from the single qubit case. An example occurs in the probability for two qubits to remain in their initial joint state. It is found to have a two-frequency beat note in the excitation-revival signals. This is absent in the single-qubit case. 

In cases of coherent-state preparation we were able to obtain convenient analytic formulas not previously available, and showed that the analytic predictions compare favorably with exact numerical results.  Expressions for individual collapse and revival signals were derived, providing formulas for the width, height and time of individual revivals.  

Tracking of entanglement evolution, as a principal measure of intrinsically quantum coherence, is complicated even in the two-qubit case because several varieties of entanglement are present. We concentrated on qubit-qubit entanglement as our primary case study, which required a trace over the oscillator's degrees of freedom. The $4 \times 4$ two-qubit density matrix yielded a compact concurrence formula that included sequences of collapse and revival signals, and also indicated a route to control over them. Analytic expressions not previously available were derived for revival strength and timing. 

In this report, we have assumed that the system dynamics is not affected by interaction with a larger environment. It is an interesting question to determine in what ways an external environment will severely or not severely impact the results presented. In order to lift this limitation, further analysis of the decoherence behavior of qubit-oscillator systems in the quasi-degenerate ultra-strong coupling regime is necessary. The possibility of generating non-classical states of the oscillator by letting it interact with multiple qubits in the quasi-degenerate ultra-strong regime is important but is not addressed in this report. \\

%===================================================
\begin{acknowledgments}
We thank S. Wallentowitz, M. Y\"{o}na\c{c}, and D. Jain for advice on some calculations.
Partial financial support was received from DARPA HR0011-09-1-0008, ARO W911NF-09-1-0385, and NSF PHY-0855701.
\end{acknowledgments}

%===================================================
%\noindent APPENDIX: CALCULATION OF $S_{1}(t)$\label{a.S_1}

\appendix*
\section{CALCULATION OF $S(t,\omega_0)$}\label{a.S_1}

The infinite sum that we want to calculate is:
\beq\label{e.A.sum}
S(t,\omega_0)=\sum_{N=0}^{\infty}P(N)\cos{(\omega_0\scpr{N_1}{N_0}t)}.
\eeq
The Poisson distribution $P(N)$ has an average and variance equal to $|\alpha|^2$. If $|\alpha|$ is big enough, one can approximate $P(N)$ to a Gaussian with the same mean and variance and justify the replacement: 
\beq\label{e.A.Gaus}
P(N) = \frac{e^{-|\alpha|^2}|\alpha|^{2N}}{N!} \to \frac{1}{\sqrt{2\pi|\alpha|^2}}e^{-\frac{(N-|\alpha|^2)^2}{2|\alpha|^2}}.
\eeq 
An analytic form for the infinite sum is desirable but challenging because of the Laguerre polynomial appearing in the definition of $\scpr{N_1}{N_0}$. One notes that if $|\alpha\beta|\ll1$, one can approximate the Laguerre polynomial by retaining only the first three terms of it to get:  
\beq\label{e.A.app_Om}
\omega_0\scpr{N_1}{N_0}\approx \omega_0 e^{-x/2}\left(1-Nx+\frac{N(N-1)}{4}x^2\right) 
\eeq
where $x=\beta^2$.
When this approximation is justified we can insert it in summation (\ref{e.A.sum}) and get:
\beqa\label{e.A.sum_app}
S(t,\omega_0) &=& Re\sum_{N=0}^{\infty}P(N)\nonumber\\
& \times & \exp{\left[i\tau\left(1-Nx+\frac{N(N-1)}{4}x^2\right)\right]},\nonumber
\eeqa
where we have defined a dimensionless scaled time by 
$$\tau=\omega_0 t e^{-x/2}.$$
Now we use the Poisson sum formula, according to which we get:
\beq\label{e.Poisson}
S(t,\omega_0)=Re\left[\sum_{k=-\infty}^{\infty}\bar{S}_{k}(t,\omega_0)+\frac{1}{2}P(0)\exp{\left(i\tau\right)}\right]
\eeq
where 
\beqa
\bar{S}_{k}(t,\omega_0)= &&\int_{0}^{\infty} \mathrm{d}n P(n)e^{2i\pi kn}\nonumber\\
&&\exp{\left[i\tau\left(1-nx+\frac{n(n-1)}{4}x^2\right)\right]}.\nonumber
\eeqa

Using the replacement  (\ref{e.A.Gaus}), we see that $\bar{S}_{k}(t,\omega_0)$ becomes a Gaussian integral, and when the excitation number of the oscillator is great enough, $|\alpha|^2 \gg 1$, one can extend the lower limit of the integral to $n = -\infty$  and evaluate the integral analytically. The result is:
\begin{align}\label{e.A.S_k}
\bar{S}_{k}(t,\omega_0)=\frac{1}{\left(1+(yf/2)^2\right)^{1/4}}\exp{(\Phi_{Re}+i\Phi_{Im})},
\end{align} 
where
\begin{align}\label{e.A.Phi_Re}
\Phi_{Re}&=\frac{|\alpha|^2}{2(1+(yf/2)^2)}\left(1-(y+yx/4-2\pi k)^2\right.\nonumber\\
&\qquad\qquad+ \left.yf(y+yx/4-2\pi k)\right)-|\alpha|^2/2,\\
\Phi_{Im}&=\frac{|\alpha|^2}{2(1+(yf/2)^2)}\left((1-(y+yx/4-2\pi k)^2)yf /2\right.\nonumber\\
&\qquad\quad\left.-2(y+yx/4-2\pi k)\right)-\theta/2-\tau,
\end{align}
and we defined:
\begin{align}
y&=\tau x,\nonumber\\
f&=|\alpha|^2x,\nonumber\\
\theta&=tan^{-1}(\pi kf)^2.
\end{align}

The contribution of $\bar{S}_{k}(t,\omega_0)$ to the sum $S(t,\omega_0)$ will be maximum when $\Phi_{Re}$ is maximum, which occurs at times around $y=2\pi k$. With this observation, and using $x \ll 1,\ f \ll 1$, we can simplify the expression for $\Phi_{Re}$ by neglecting terms that are of the order of $yx$, $(y- 2\pi k)f^2$, $f^3$, $(y-2\pi k)^2f$ and higher powers of these terms to get:
\begin{align}\label{e.A.Phi_re_app}
\Phi_{Re}&=-\frac{|\alpha|^2}{2\left(1+(\pi kf)^2\right)}\left(y-2\pi k(1+f/2)\right)^2.
\end{align}
Similarly, one can simplify the expression for $\Phi_{Im}$ to get:
\begin{align}\label{e.A.Phi_im_app}
\Phi_{Im}=-\frac{tan^{-1}{(\pi kf)^2}}{2}+\frac{1}{x}\left(y(1+f)-2\pi k f\right).
\end{align} 
Similarly, one can approximate the prefactor in (\ref{e.A.S_k}) to be:
\begin{align}
\frac{1}{1+(yf/2)^2}\approx\frac{1}{1+(\pi k f)^2}.
\end{align}
Thus, $\bar{S}_k(t,\omega_0)$ takes the simplified form:
\beqa\label{e.A.dy_S_k}
\bar{S}_{k}(t,\omega_0)&=&exp{\left(\frac{-(\tau-\tau_k)^2|\alpha|^2x^2}{2\left(1+(\pi kf\right)^2)}+i\Phi_{Im}\right)}\nonumber\\
&\times&\frac{1}{\left(1+(\pi kf)^2\right)^{1/4}},
\eeqa
where $\Phi_{Im}$ is given by (\ref{e.A.Phi_im_app}) and $\tau_k$ is defined as
\beq
\tau_k=2\pi k(1+f/2)/x.
\eeq

We now note that the amplitude of $\bar{S}_k(t,\omega_0)$ is much greater than $P(0)=e^{-|\alpha|^2}$. Thus, we can neglect the term $\frac{1}{2}P(0)\exp{\left(i\tau\right)}$ from (\ref{e.Poisson}). We also note from (\ref{e.A.Phi_re_app}), that for positive time $\tau$, we must have $k\geq0$. Thus the expression for $S(t,\omega_0)$ becomes:
\beq
S(t,\omega_0)=Re\left[\sum_{k=0}^{\infty}\bar{S}_{k}(t,\omega_0)\right].
\eeq


\begin{thebibliography}{99}


\bibitem{Rabi}
I.I. Rabi, Phys. Rev. \textbf{49}, 324 (1936); \textbf{51}, 652 (1937).

\bibitem{JC}
E.T. Jaynes and F.W. Cummings, Proc. IEEE \textbf{51},  89 (1963).

\bibitem{Allen-Eberly}
L. Allen and J.H. Eberly in \textit{Optical Resonance and Two-Level Atoms} (Dover Publications, 1987).

\bibitem{Holstein}
T. Holstein, Ann. Phys. (N.Y.) \textbf{8}, 325 (1959). 

\bibitem{Irish-Schwab}
E.K. Irish and K. Schwab, \prb{68}, 155311 (2003).

\bibitem{Schwab-Roukes}
K.C. Schwab and M.L. Roukes, Phys. Today \textbf{58}, 36 (2005).

\bibitem{Blais-etal}
A. Blais, R.S. Huang, A. Wallraff, S.M. Girvin, 
and R.J. Schoelkopf, \pra{69}, 062320 (2004).

\bibitem{Wallraff-etal}
A. Wallraff, D.I. Schuster, A. Blais, L. Frunzio, R.-S. Huang,
J. Majer, S. Kumar, S.M. Girvin, and R. J. Schoelkopf, Nature
(London) \textbf{431}, 162 (2004).

\bibitem{Chiorescu-etal}
I. Chiorescu, P. Bertet, K. Semba, Y. Nakamura, C.J.P.M. Harmans,
and J.E. Mooij, Nature (London) \textbf{431}, 159 (2004).

\bibitem{Johansson-etal}
J. Johansson, S. Saito, T. Meno, H. Nakano, M. Ueda, K. Semba,
and H. Takayanagi, \prl{96}, 127006 (2006).

\bibitem{Dicke}
R.H. Dicke, Phys. Rev. \textbf{93}, 99 (1954). In this paper, the field is treated classically. 

\bibitem{Tavis-Cummings-1}
M. Tavis and F.W. Cummings, Phys. Rev. \textbf{170}, 379 (1968).

\bibitem{Tavis-Cummings-2}
M. Tavis and F.W. Cummings, Phys. Rev. \textbf{188}, 692 (1969).

\bibitem{Leek}
P. J. Leek, S. Filipp, P. Maurer, M. Baur, R. Bianchetti, J. M. Fink, M. G\"{o}ppl, L. Steffen, and A. Wallraff, \prb{79}, 180511(R) (2009).

\bibitem{Wootters}
W.K. Wootters, \prl{80}, 2245 (1998).

\bibitem{Peres}
A. Peres, \prl{77}, 1413, (1996).

\bibitem{Grobe}
R. Grobe, K. Rz\c{a}zewski and J.H. Eberly, J. Phys. B \textbf{27}, L503 (1994).

\bibitem{Niemczyk-etal}
T. Niemczyk et al., Nat. Phys. \textbf{6}, 772 (2010).

\bibitem{Forn-etal}
P. Forn-D\'iaz, J. Lisenfeld, D. Marcos, J.J. Garc\'ia-Ripoll,
E. Solano, C.J.P.M. Harmans, and J.E. Mooij, Phys. Rev.
Lett. \textbf{105}, 237001 (2010).

\bibitem{Fedorov-etal}
A. Fedorov, A.K. Feofanov, P. Macha, P. Forn-D\'iaz, C.J.P.M.
Harmans, and J.E. Mooij, Phys. Rev. Lett. \textbf{105}, 060503 (2010).


\bibitem{Irish-05}
E.K. Irish, J. Gea-Banacloche, I. Martin, and K.C.
Schwab, Phys. Rev. B \textbf{72}, 195410 (2005). 

\bibitem{Irish-07}
E.K. Irish, \prl{99}, 173601 (2007).

\bibitem{Devoret}
M. Devoret, S. Girvin, and R. Schoelkopf, Ann. Phys. (Leipzig) \textbf{16}, 767 (2007).

\bibitem{Bourassa}
J. Bourassa, J. M. Gambetta, A. A. Abdumalikov, O. Astafiev, Y. Nakamura, and A. Blais, Phys. Rev. A \textbf{80}, 032109 (2009).

\bibitem{Ashhab}
S. Ashhab and F. Nori, Phys. Rev. A 81, 042311 (2010).

\bibitem{Hausinger-10}
J. Hausinger and M. Grifoni, \pra{82}, 062320 (2010). 

\bibitem{Casanova}
J. Casanova, G. Romero, I. Lizuain, J.J. Garc\'ia-Ripoll, and E. Solano, Phys. Rev. Lett. \textbf{105}, 263603 (2010).

\bibitem{Schweber-67}
S. Schweber, Ann. Phys. (N.Y.) \textbf{41}, 205 (1967).

\bibitem{vanVleck}
J.H. Van Vleck, Phys. Rev. 33, 467 (1929). 

\bibitem{Shirley-65}
J. Shirley, Phys. Rev. 138, B979 (1965).
    
\bibitem{Leggett}
A.J. Leggett, S. Chakravarty, A.T. Dorsey, M.P.A. Fisher, A. Garg, and W. Zwerger, Rev. Mod. Phys. \textbf{59}, 1 (1987).

\bibitem{Sandu}
T. Sandu, Phys. Lett. A \textbf{373}, 2753 (2009).

\bibitem{Collapse-Revival}
J.H. Eberly, N.B. Narozhny, and J.J. Sanchez-Mondragon, \prl{44}, 1323 (1980); 
N.B. Narozhny, J.J. Sanchez-Mondragon, and J.H. Eberly, \pra{23}, 236 (1981); 
H.I. Yoo, J.J. Sanchez-Mondragon, and J.H. Eberly, \jpa{14}, 1383 (1981).

\bibitem{Abramowitz}
M. Abramowitz and I.A. Stegun in \textit{Handbook of Mathematical Functions}, 13.5.14 on Pg. 508  (Tenth Printing, 1972).

\bibitem{Deng}
Z. Deng, Opt. Comm., \textbf{54}, 222 (1985).

\bibitem{Lee-06}
J. Lee, M. Paternostro, M.S. Kim, and S. Bose, Phys Rev. Lett. \textbf{96}, 080501 (2006).

\bibitem{Yonac-10}
M. M. Y\"{o}na\c{c} and J.H. Eberly, Phys. Rev. A \textbf{82}, 022321 (2010).

\bibitem{Tessier}
T. E. Tessier, I.H. Deutsch, A. Delgado, and I. Fuentes-Guridi, \pra{68}, 062316 (2003).

\bibitem{Jing-Ficek}
J. Jing, Z.G. L\"u, and Z. Ficek, \pra{79}, 044305 (2009).

\bibitem{Ficek_etal}
Z. Ficek, J. Jing, and Z.G. L\"u, Phys. Scr. \textbf{T140}, 014005 (2010).

\bibitem{Chen_etal}
Q.H. Chen, Y. Yang, T. Liu, and K.L. Wang, \pra{82}, 052306 (2010). 

\bibitem{Leon-Sabin}
J. Le\'{o}n and C. Sab\'{i}n, Phys. Rev. A \textbf{79}, 012304 (2009).

\bibitem{X-Matrix}
T. Yu and J.H. Eberly, Quantum Information and Computation \textbf{7}, 459 (2007).

\bibitem{Horodecki}
R. Horodecki, P. Horodecki, M. Horodecki, and K. Horodecki, Rev. Mod. Phys. \textbf{81}, 865Ð942 (2009).


\end{thebibliography}
\end{document}